\documentclass{jpsj3}
\usepackage{txfonts}

\title{Origins of Valley Current Reversal in Partially Overlapped Graphene Layers} 
\author{Ryo Tamura}
\inst{Faculty of Engineering, Shizuoka University, 3-5-1 Johoku, Hamamatsu 432-8561, Japan}

\abst{Using the tight-binding model,
 we investigate the valley current
of the `low-bi-up'
and `low-bi-low' graphene junction,
where `low' and `up'
are respectively the lower and upper graphene layers
extended from the
central AB stacking bilayer graphene layer, `bi'.
Source and drain electrodes connect with 
the left and right monolayer regions, respectively,
and thus the total current is forced to
flow through the interlayer path in the low-bi-up junction.
We measure valley current reversal (VCR) using
the average of $\frac{1}{2} \sum_{\nu =\pm}(T_{\nu,-\nu}-T_{\nu,\nu})$
per lateral wave number, where $T_{\nu,\nu'}$ denotes the electron
transmission rate 
from the left $K_{\nu'}$ valley to the right $K_{\nu}$ valley.
Without the vertical electric field, the VCR is less than half in both junctions.
This VCR is attributed to monolayer--bilayer matching.
As the vertical field intensifies,
the VCR
declines 
in the low-bi-low junction, but
increases 
to about 0.8 in the low-bi-up junction.
This VCR enhancement originates from interlayer matching.
Analytic scattering matrixes
elucidate these matching effects.
Experiments of VCR detection are also proposed.}

\begin{document}
\maketitle

\section{Introduction} 

The discovery of the exfoliation synthesis\cite{exfoil} has opened up avenues to Hall effects \cite{QHE,QHE2} of the ultimate thin layer, graphene (G) \cite{(1),graphene-review,graphene-review2}.
The concept has been generalized to spin (valley) 
Hall conductivity $\sigma_{\rm s}$ ($\sigma_{\rm v}$)
\cite{r9.,monolayer-theory2}. Here, the valleys refer to the inequivalent
corner points in the Brillouin zone 
and are denoted by $K_+$ and $K_-$
on the analogy of up ($+$) and down ($-$) spins.
The $\sigma_{\rm s}$ ($\sigma_{\rm v}$) equals $(J_+-J_-)/V$,
where $J_+$ and $J_-$ denote contributions to the charge current
from the spin $+$ and $-$ (valley $K_+$ and $K_-$).
The transverse voltage $V$ induces the longitudinal spin (valley) current $J_+-J_-$, and vice versa.
Compared with the charge current $J_++J_-$, it is not easy to detect the $J_+-J_-$, and the nonlocal resistance $R_{nl}$ is an alternative.
The $R_{nl}$ enhanced by the magnetic field was attributed
to $\sigma_{\rm s}$ in Refs. \cite{s1.} and  \cite{s2.} , 
but the insensitivity to the in-plane magnetic field suggests 
 the valley as the origin \cite{s3.,(3)}.
This closely connects with valleytronics, where valleys carry, store, and manipulate information \cite{valley-review,valley-review2,valley-review3},
 in the analogy to spintronics \cite{s32.,spin}.
The dissipationless `pure' valley current (VC) with zero charge current is particularly appealing.
This pure VC is a likely origin of the giant $R_{nl}$ in the bilayer G \cite{bilayer,bilayer2,bilayer3,recent-experiment} and the monolayer G\cite{bilayer-theory-strain,monolayer-theory2, monolayer,monolayer2,monolayer-theory}.
Quantum pumping can also generate the pure VC \cite{34.}.

In addition to $\sigma_{\rm v}$, various proposals on the G-based valleytronics exist.
The line defect \cite{disclination-pair,disclination-pair2,disclination-pair3,disclination-pair4}, 
strain field \cite{bubble,bubble2,bubble3,37.,38.}, 
zigzag edge states \cite{41.,a4-3.,a4.,a4-2.}, and twisted bilayer G \cite{20.} work as the VC filters that transmit only one of $J_+$ or  $J_-$.
The $J_+$ stream branches off from the $J_-$ streams through strain fields \cite{4,a7.}, magnetic-electric barriers \cite{36.}, transistor interfaces \cite{5}, 
and spatially alternating vertical electric fields 
\cite{top-back-gate}.
A superconducting contact \cite{25.,21.} and the splitting of Landau levels \cite{19.} enable  the detection of  the valley polarization.
There also exist theoretical proposals for optical generation 
\cite{a1.,a2.,L-1.,L-5.,L-6.,L-7.,Dixit-1,Dixit-2,L-3.,L-10.} and detection \cite{L-8.,L-9.}.
In these discussions, the intervalley scattering disturbs the VC randomly
and merely produces noise.
However, the controlled intervalley scattering 
enables us to detect 
the VC reversal (VCR) by the $R_{nl}$ sign change.
The VCR is a crucial function in valleytronics, for example, as a `not' logic gate
in pure VC. The unidirectional propagation of the zigzag edge state in a given valley becomes opposite when the energy moves across the neutral Fermi level. \cite{41.,a4.}.
It explains the VCR in the $p$-$n$ junction of the zigzag G ribbon \cite{40.}.
The zone folding illustrates the VCR 
origin of superlattice graphene (SG) sandwiched between pristine G regions 
\cite{graphene-precession,graphene-precession2,graphene-precession3}.
Compared with these VCRs, the VCR origin remains unclear in 
the partially overlapped G (po-G);
 Li et al. argued that Fano resonance suppresses the intravalley transmission
but did not show what enhances the intervalley transmission \cite{gra-gra-junction}.
Their numerical outputs are partially inconsistent with the Fano resonance.
The interlayer potential difference improves the VCR in the side contacted
armchair nanotubes \cite{tamura-2021}, but it is not discussed in Ref. \cite{gra-gra-junction} . 
The VCR is outside the scope of other theoretical works about the po-G
\cite{7.,9.,10.,12.,13.,14.,16.,17.,a3.,(4)} and the bilayer--monolayer interface \cite{15.,18.}.

$\sigma_{\rm v}$ and $R_{nl}$ have been calculated using the semiclassical formulation (SCF) \cite{(5)} and Landauer--B\"{u}ttiker formulation (LBF). 
\cite{2.,3.,4.,5.,6.,42.}
The SCF leads to the scaling relation $R_{nl} \propto \sigma_{\rm v}^2\rho^3$ with the Ohmic resistivity $\rho$ \cite{valley-review2,bilayer,(6)}.
The cubic law $R_{nl} \propto \rho^{3}$ is  $\sigma_{\rm v}$ evidence, 
but the absence of direct $\sigma_{\rm v}$ detection causes uncertainty \cite{5.,6.}.
Nonzero $\sigma_{\rm v}$ originates from
either bulk states \cite{s34.} or topological edge states in the SCF \cite{(2),(7)}, whereas nontopological edge transport also satisfies
the $\rho^{3}$ scaling law in the LBF \cite{Roche-1}.
There also exist non-VC pictures of $R_{nl}$:
electron fluid viscosity \cite{r25.,s13.,s37.,s39.}, edge charge accumulation \cite{8.}, and the Nernst-Ettingshusen effect \cite{s21.,s22.}. 
In this paper, we perform an LBF calculation of the VCR that is a probe for the bulk VC contribution to the $R_{nl}$. 
Although the electron correlation causes superconductivity, 
 the critical temperature is considerably lower
than the typical temperature  in $R_{nl}$ measurements \cite{(8),(9),(10)}.
The spin splitting 
of the low-energy conduction and valence bands is lower
than 0.1 meV for the vertical electric field
 in this paper \cite{a11.,a12.}.
Accordingly, we exclude these two aspects from the scope of this paper.

The rest of this  paper is organized as follows.
In Sect. 2, we present the tight-binding (TB) models and define
 the VCR indicator $\tilde{g}_{\rm v}$.
The perfect pure VCR is realized when $\tilde{g}_{\rm v}$ reaches one. 
In Sect. 3, we derive analytic formulas of
the VCR transmission rate about the normal incidence.
In Sect. 4, we  compare these analytical results with the exact numerical data and 
 clarify the two kinds of VCR origins. 
The wave function analysis presents intuitive pictures of the VCR.
In Sect. 5,  we propose experimental 
probes for the bulk VC contribution to  $R_{nl}$.
In Sect. 6, we present a summary and the conclusions.

\section{TB Models and  VCR Transmission Rates} 
\begin{figure}
\begin{center}
\includegraphics[width=0.7\linewidth]{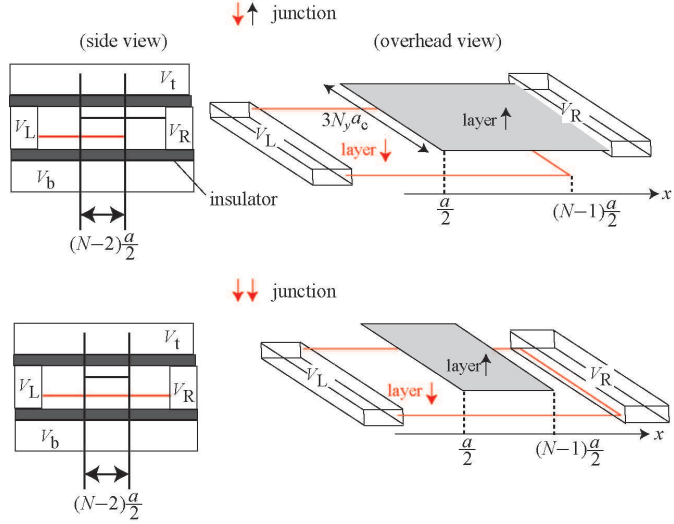}
\caption{(Color online) Arrangements of electrodes and graphene layers $\downarrow$ and $\uparrow$.
 $V_{\rm t}$ and $V_{\rm b}$ denote
the top and bottom gate electrodes, respectively.
 $V_{\rm L}$ and $V_{\rm R}$ denote
the bias electrodes.
}
\label{fig-intro}
\end{center}
\end{figure}

\begin{figure}
\begin{center}
\includegraphics[width=0.5\linewidth]{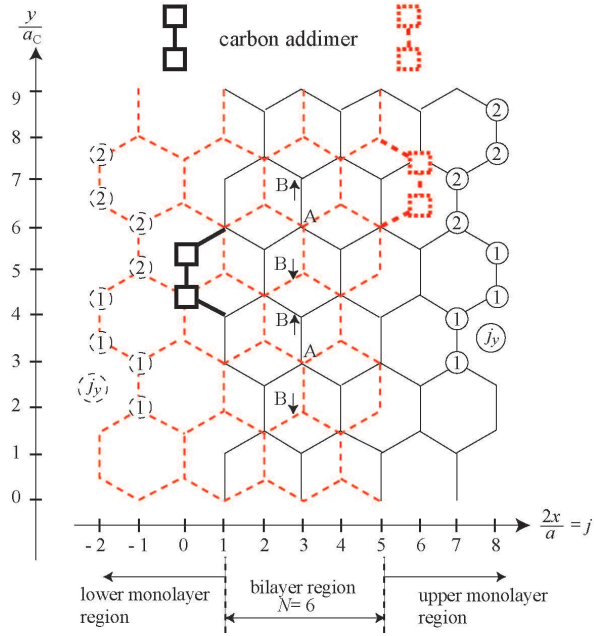}
\caption{(Color online) Indexes $j,j_y, A,$ and $ B$ 
and atomic positions in the $\downarrow\uparrow$ junction.
Layers $\downarrow$ and $\uparrow$ are parallel to the $xy$ plane.
}\label{fig-intro-xy}

\end{center}
\end{figure}

We consider the two kinds of po-G, $\downarrow\uparrow$
and $\downarrow\downarrow$, as shown in Fig. \ref{fig-intro},  
where $\downarrow$ and $\uparrow$ denote
the lower and upper graphene layers, respectively.
The top $V_{\rm t}$ and bottom $V_{\rm b}$ gate electrodes
exert a vertical electric field on the two layers.
The bilayer region is limited spatially and each of the bias electrodes
$V_{\rm L}$ and $V_{\rm R}$
connects with only one of the layers.
In the $\downarrow\uparrow$ junction, 
$V_{\rm L}$ and $V_{\rm R}$ 
connect with
the $\downarrow$ and $\uparrow$ layers, respectively.
In the $\downarrow\downarrow$ junction, 
both $V_{\rm L}$ and $V_{\rm R}$ 
connect with
the $\downarrow$ layer.
Figure \ref{fig-intro-xy} 
 shows the atomic structures of
the $\downarrow\uparrow$ junction in the case of $N=6$.
The dotted and solid lines 
represent $\downarrow$ 
and $\uparrow$ layers, respectively.
Integer indexes ($j$, $j_y$)
and sublattice indexes $(A,B)$ specify the atomic coordinates $(x,y)$ as
$x=\frac{a}{2}j$,
$y_{\rm A,\downarrow}=y_{\rm A,\uparrow}=3 a_{\rm c} [j_y+ \frac{1+(-1)^j}{4}]$, $y_{\rm B,\downarrow}=y_{\rm A}-a_{\rm c}$,
 and $y_{\rm B,\uparrow}=y_{\rm A}+a_{\rm c}$
with the lattice constant $a$
and the bond length $a_{\rm c}=a/\sqrt{3}$. 
The bilayer region is limited in the range $1\leq j \leq N-1$
with the geometrical overlap length $(N-2)a/2$.
The left $j \leq 0$ 
and right $N \leq j$ monolayer regions respectively 
correspond to
the $\downarrow$ and $\xi$ layers in the $\downarrow\xi$ junction
($\xi= \downarrow, \uparrow)$. 
Bonded squares denote carbon dimers added to
armchair edges.
The addimer $(x,y)$ positions are represented by
$(0, (3m+2)a_{\rm c})$ and $(Na/2, (3m'+1)a_{\rm c})$ with integers $m$ and $m'$.
When not explicitly noted, we assume the
perfect armchair edges. 

The wave functions at sublattices A and B are represented by
$A_{\xi,j,j_y}$ and $B_{\xi,j,j_y}$
with the layer index $\xi= \downarrow, \uparrow$.
Using the periodic boundary condition 
$(A,B)_{j_y+N_y}=(A,B)_{j_y}$ with the transverse width 
$3N_ya_{\rm c}$, we can reduce the dimension of 
the TB equations of the bilayer region 
as 
\begin{eqnarray}
E\vec{c}_{\downarrow,j}
&=& 
(h_0-\varepsilon)\vec{c}_{\downarrow,j}+h_1\left(\vec{c}_{\downarrow,j-1}+
\vec{c}_{\downarrow,j+1}\right)
\nonumber\\
&& 
+W_0\vec{c}_{\uparrow,j}
+W_1\left(\vec{c}_{\uparrow,j-1}+
\vec{c}_{\uparrow,j+1}
\right),
\label{TB-down}
\end{eqnarray}
\begin{eqnarray}
E\vec{c}_{\uparrow,j}&=& 
(h_0+\varepsilon) \vec{c}_{\uparrow,j}
+h_1^*\left(\vec{c}_{\uparrow,j-1}+
\vec{c}_{\uparrow,j+1}\right)
\nonumber\\
&& 
+W_0^*\vec{c}_{\downarrow,j}
+W_1^*\left(\vec{c}_{\downarrow,j-1}+
\vec{c}_{\downarrow,j+1}
\right),
\label{TB-up}
\end{eqnarray}
where 
\begin{equation}
\;^t\vec{c}_{\xi,j}=e^{-\frac{i}{2}\kappa\left[
(-1)^j+1 \right]} (A_{\xi,j,0}, B_{\xi,j,0}),
\end{equation}
\begin{equation}
h_0 =
\gamma_0\left(
\begin{array}{cc}
0 & 1 \\
1 & 0
\end{array}
\right),\;\;
h_1=
\gamma_0\left(
\begin{array}{cc}
0 & e^{i\kappa}
\\
e^{-i\kappa} & 0
\end{array}
\right),
\end{equation}

\begin{equation}
W_0= \left(
\begin{array}{cc}
\gamma_1 & \gamma_4 \\
\gamma_4 & 
\gamma_3e^{-2i\kappa}
\end{array}
\right),\;
W_1= 
e^{-i\kappa} \left(
\begin{array}{cc}
0 & \gamma_4
\\
\gamma_4 & \gamma_3
\end{array}
\right),
\end{equation}
$\kappa= \frac{\sqrt{3}}{2}k_ya$, and $k_y$ is the $y$ component of the wave number vector.
The vertical electric field induces the interlayer difference $2\varepsilon$ in site energy \cite{note}.
Two kinds of TB models, $\gamma_1\gamma_3\gamma_4$-TB
and $\gamma_1$-TB models,
are used, where
the latter is an approximation of the former.
According to Ref. \cite{TB-parameter} , 
the $\gamma_1\gamma_3\gamma_4$-TB parameters 
are $\gamma_0=-3.12$ eV , $\gamma_1=0.377$ eV, $\gamma_3=0.29$ eV, and 
$\gamma_4=0.12$ eV.
The $\gamma_1$-TB parameters are the same as the $\gamma_1\gamma_3\gamma_4$-TB parameters except that $\gamma_3=\gamma_4=0$.
The Hamiltonian elements of the addimers are defined in the same way.

Applying the exact method in Ref. \cite{tamura-2019} to Eqs. (\ref{TB-down}) and (\ref{TB-up}), we can calculate $T_{\nu',\nu}(\kappa)$ that denotes the transmission rate
from the left $K_\nu$ state to the right $K_{\nu'}$ state.
The Landauer's formula conductivity $g$ (conductance per width $3a_{\rm c}$
in the unit of 2$e^2/h$ ) 
and the {\it absolute} VCR conductivity $g_{\rm v}$ are represented by
\begin{equation}
\left(
\begin{array}{l}
g
\\
g_{\rm v}
\end{array}
\right)
=\frac{1}{N_y} 
\sum_{m=-M}^M 
\left(
\begin{array}{l}
g'(m \Delta \kappa )
\\
g_{\rm v}' (m \Delta \kappa )
\end{array}
\right),
\label{g-gvr}
\end{equation}
where 
\begin{equation}
\left(
\begin{array}{l}
g'(\kappa)
\\
g_{\rm v}'(\kappa)
\end{array}
\right)
=\sum_{\nu=\pm}\sum_{\nu'=\pm}
\left(
\begin{array}{l}
1
\\
\frac{-\nu\nu'}{2}
\end{array}
\right)
T_{\nu',\nu}(\kappa).
\end{equation}
The {\it relative} VCR conductivity is represented by $g_{\rm v}/g$.
Owing to the periodic boundary condition
with the period $3N_ya_{\rm c}$,
 the transverse wave number $k_y$ is discrete 
as $k_y=m\frac{2\pi}{3N_ya_{\rm c}}$ with integers $m$.
As $\kappa= \frac{\sqrt{3}}{2}k_ya$, 
the interval of the discrete $\kappa$ is represented by
$\Delta \kappa = \frac{\sqrt{3}}{2}a \Delta k_y = \frac{\pi}{N_y}$. 
The wave number vector $\vec{k}=(k_x,k_y)$ 
of the monolayer region
must satisfy 
the dispersion relation
\begin{equation}
\frac{(E\pm\varepsilon)^2}{\gamma_0^2}=\left(2\cos\left(\frac{a}{2}k_x\right)
+\cos\kappa\right)^2+\sin^2\kappa
\label{monolayer-gap}
\end{equation}
indicating the energy gap $|E \pm \varepsilon| \leq |\gamma_0 \sin \kappa|$ 
in a subband with a fixed $\kappa$.
 The integer $M$ in Eq. (\ref{g-gvr})
corresponds to the maximum of the allowed $|k_y|$
and is represented by
\begin{equation}
M=
\left\{
\begin{array}{l}
{\rm Int} \left[\frac{N_y}{\pi}\arcsin\left|\frac{|E|-|\varepsilon|}{\gamma_0} \right|\right]
\;\;\cdots(\downarrow\uparrow \;{\rm junction})
\\
{\rm Int} \left[\frac{N_y}{\pi}\arcsin\left|\frac{E+\varepsilon}{\gamma_0} \right|\right]
\;\;\cdots(\downarrow\downarrow \;{\rm junction})
\end{array}
\right.,
\label{M-condition}
\end{equation}
where Int$[x]$ denotes the maximum integer that does not exceed $x$. 
The average VCR transmission rate
for the active $2M+1$ subbands is represented by
\begin{equation}
\widetilde{g}_{\rm v} =\frac{N_y}{2M+1} g_{\rm v}
\label{ave-gv}
\end{equation}
that is relevant to both $g_{\rm v}$ and $g_{\rm v}/g$.
When $\widetilde{g}_{\rm v}=1$,
the VCR transmission rates become perfect 
for both $T_{+,-}$ and $T_{-,+}$ in all the $2M+1$ subbands.
It follows that $g_{\rm v}$ and $g_{\rm v}/g$ 
concurrently reach their upper limits, $(2M+1)/N_y$
and $\frac{1}{2}$.
Under the condition $\tilde{g}_{\rm v}=1$,
the pure VC changes into the inverse pure VC without losing
its intensity. The condition $T_{+,-}=T_{-,+}$,  which follows 
the condition $\widetilde{g}_{\rm v}=1$, is 
necessary for this pure VCR.
When only one of $T_{+,-}$ or $T_{-,+}$ reaches one, a pure VC changes
into a nonpure VC.
Even when the relative VCR conductivity is perfect ($g_{\rm v}/g=\frac{1}{2}$),
the absolute VCR conductivity may be far from the upper limit
($ g_{\rm v} \ll (2M+1)/N_y$).
Inversely,
$g_{\rm v}$ may be large with a small $g_{\rm v}/g$ owing to a large $M/N_y$.
The condition $\widetilde{g}_{\rm v} \simeq 1$ rules out
these cases where only one of $g_{\rm v}/g$ or $g_{\rm v}$ is large.
Refer to Appendix A for the detailed VC formulas.

The signs of $\gamma_0$ and
$\gamma_{4}$ are reversed compared with those shown in Ref. \cite{TB-parameter} by the transformation $(A',B')=(A,-B)$. 
The transformation $
(A'_{\downarrow},B'_\downarrow)=(A_{\downarrow},-B_\downarrow), $ $(A'_{\uparrow},B'_\uparrow)=(-A_{\uparrow},B_\uparrow)$ proves that
the $\gamma_1$-TB model
has the symmetry
\begin{equation}
\tilde{g}_{\rm v}(E,\varepsilon)=\tilde{g}_{\rm v}(-E,-\varepsilon)
\label{sym2}
\end{equation}
for both the $\downarrow\uparrow$ and $\downarrow\downarrow$ 
junctions.
Even in the $\gamma_1\gamma_3\gamma_4$-TB model, Eq. (\ref{sym2})
holds approximately. Refer to Ref. \cite{tamura-2019} for  this insignificance of $\gamma_3$ and $\gamma_4$.
Appendix A proves that the $\downarrow\uparrow$
junction satisfies
\begin{equation}
\tilde{g}_{\rm v}(E,\varepsilon)=\tilde{g}_{\rm v}(E,-\varepsilon)
\label{sym1}
\end{equation}
in both the $\gamma_1$-TB and $\gamma_1\gamma_3\gamma_4$-TB models,
whereas Eq. (\ref{sym1}) is invalid for the $\downarrow\downarrow$ junction.
Owing to Eq. (\ref{sym2}) that holds approximately (exactly) in the $\gamma_1\gamma_3\gamma_4$-TB ($\gamma_1$-TB) model, 
we eliminate the negative $\varepsilon$ from the rest of this paper.

\begin{table}
\begin{tabular}{c|c|c|c|c}
&
\multicolumn{2}{c|}{$\frac{dE}{dk} > 0\;\; (p=+)$} &
\multicolumn{2}{c}{$\frac{dE}{dk} < 0\;\; (p=-) $} \\
\hline
&$K_+$&$K_-$& $K_+$& $K_-$\\ 
\hline
bilayer & $k_l^{(+)}$ & $-k_l^{(-)}$ 
& $k_l^{(-)}$ & $-k_l^{(+)}$ 
\\
$l=+,-$ & & & & 
\\ \hline
monolayer & $k_\xi^{(+)}$ & $-k_\xi^{(-)}$ 
& $k_\xi^{(-)}$ & $-k_\xi^{(+)}$ 
\\
$\xi=\downarrow, \uparrow$ & & & &
\\ \hline
& \multicolumn{4}{c}{ $K_+$}
\\ \hline
$E >0$ &\multicolumn{4}{c}{$k_+^{(\sigma)}\simeq k^{(\sigma)}_\downarrow,
\;\; k_-^{(\sigma)}\simeq k^{(\sigma)}_\uparrow $ }
\\ \hline
$E <0$ &\multicolumn{4}{c}{$k_+^{(\sigma)}\simeq k^{(\sigma)}_\uparrow,
\;\; k_-^{(\sigma)}\simeq k^{(\sigma)}_\downarrow $ } \\
\end{tabular} 
\caption{ Relation of the indexes $l,\sigma$
to the layers $\xi= \downarrow,\uparrow$,  the sign of the group velocity $
p=\frac{dE}{dk}/\left|\frac{dE}{dk}\right|$, and the valleys $K_{\pm}$.
}
\end{table}

\section{$\gamma_1$-TB Calculation of  Zero Lateral Wave Number }
\begin{figure}
\begin{center}
\includegraphics[width=0.7\linewidth]{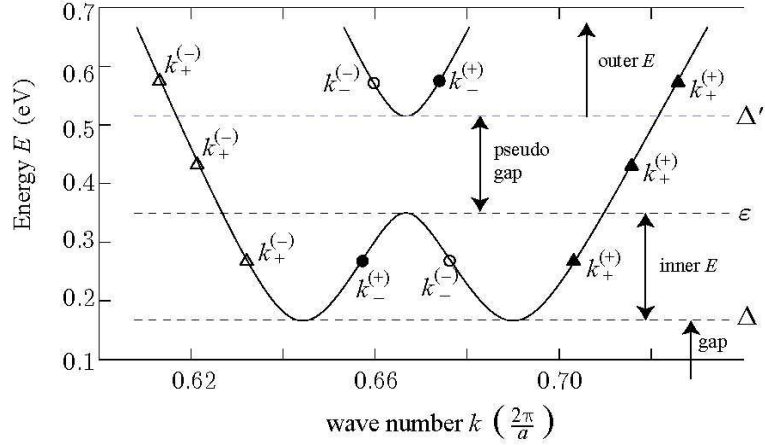}
\caption{Dispersion relation of the bilayer region 
calculated using the $\gamma_1$-TB model for $\varepsilon=$ 0.35 eV
and $\kappa=0$.
}\label{fig-band}
\end{center}
\end{figure}

In Sect. 3, we discuss the $\gamma_1$-TB
calculation of subband $\kappa=0$.
Figure \ref{fig-band} exemplifies the wave number $k_l^{(\sigma)}$ at the $K_+$ valley 
with sign indexes $l=\pm$ and $\sigma=\pm$.
At the other valley $K_-$, $k = -k^{(\sigma)}_l\simeq -4\pi/(3a)$.
The group velocity $\frac{dE}{dk}$ at the $K_+$ valley
is positive (negative) when $\sigma=+$ $(\sigma=-)$.
For the positive (negative) group velocity states,
the index $\sigma$ corresponds to the
$K_\sigma$ ($K_{-\sigma})$ valley.
The index $l$ is assigned according to the condition 
$\left|k^{(\sigma)}_+-4\pi/(3a) \right| \geq \left|k^{(\sigma)}_--4\pi/(3a)\right|$.
When $lE$ is positive (negative),
the $k_l$ dispersion line is similar to the monolayer $\downarrow$
($\uparrow$) dispersion line.
In this sense, the index $l$ is the roots layer index.
Table I summarizes the physical meaning of the indexes $l$ and $\sigma$.
Figure \ref{fig-band} also demonstrates the bilayer propagating mode number $N_{\downarrow\uparrow}$
at each valley 
as
\begin{equation}
N_{\downarrow\uparrow}
=
\left\{
\begin{array}{l}
0 \;\;\; \cdots |E| < \Delta\;\;\; ({\rm gap)}
\\
4 \;\;\; \cdots \Delta < |E| < \varepsilon\;\;\; ( {\rm inner \;\; } E)
\\
2 \;\;\; \cdots \varepsilon < |E| < \Delta'
\;\;\; ({\rm pseudogap)}
\\
4 \;\;\; \cdots \Delta' < |E| \;\;\; ({\rm outer \;\; } E)
\end{array}
\right.
\end{equation}
with the gap edge 
$\Delta =\frac{\gamma_1 \varepsilon}{\sqrt{4\varepsilon^2+\gamma_1^2}}$
and the upper pseudogap edge $\Delta' =\sqrt{\varepsilon^2+\gamma_1^2}$ \cite{a10.,a10-2.,a10.tuika,a10-note.} . 

The gap and pseudogap regions are
excluded in Sect. 3, as their smaller mode number suppresses 
the conductivity.
In Sect. 3.1, we explain the exact formulas of the inner and outer $E$
regions.
Refer to Ref. \cite{tamura-2019} for the exact calculation method of the gap and pseudogap regions.
In Sects. 3.2 and 3.3, $k_l^{(\sigma)}$ approximates $\frac{4\pi}{3a}$ 
in the scattering between the monolayer and bilayer regions.
In Sect. 3.3, we neglect the difference between
$\left|k_l^{(+)}-\frac{4\pi}{3a}\right|$ and
$\left|k_l^{(-)}-\frac{4\pi}{3a}\right|$
in the propagation through the bilayer region.
The transmission rate is concisely expressed in Sect. 3.3 when $N$ is a multiple of three. 
This expression is interpreted according to
the wave function nature.
Except when $\downarrow\downarrow$ is explicitly referred to,
 in Sect. 3, we mainly discuss the $\downarrow\uparrow$ junction.
It is straightforward to derive the $\downarrow\downarrow$ formulas
from the $\downarrow\uparrow$ formulas.

\subsection{Exact calculations}
\begin{figure}
\begin{center}
\includegraphics[width=0.5\linewidth]{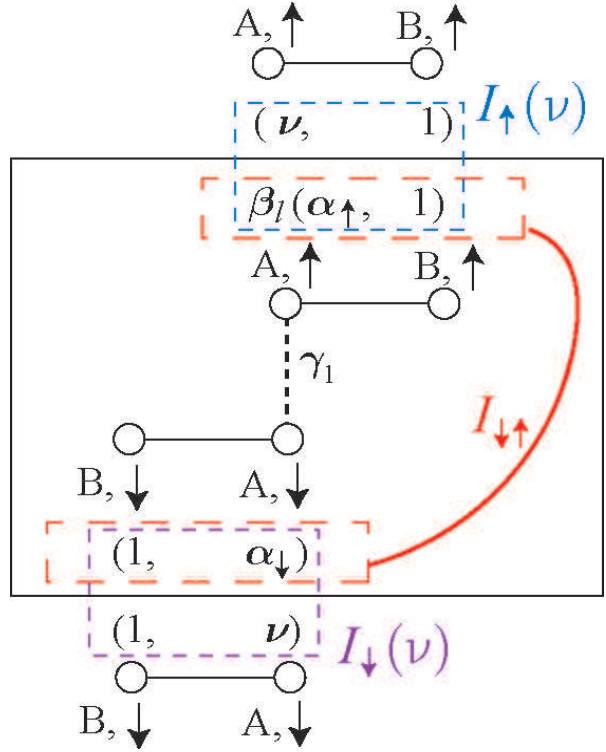}
\caption{(Color online) Cross-sectional view of elements
of Eqs. (\ref{wf-D}), (\ref{wf-mu}), and (\ref{wf-mu-up}),
where $\alpha_\downarrow =\sigma l \alpha_l$ and $\alpha_\uparrow= \alpha_\downarrow(E-\varepsilon)/(E+\varepsilon)$. 
Refer to the main text about the interlayer matching $I_{\downarrow\uparrow}$
and monolayer--bilayer matching $I_\xi(\nu)$ where $\xi =\downarrow,\uparrow$.
}\label{fig-wavefunction}
\end{center}
\end{figure}

The dispersion relation and wave function of the bilayer region are represented by
\begin{equation}
\cos
\left(\frac{a}{2}k^{(\sigma)}_{l}\right) 
=\frac{-1}{2}\left(1+l \sigma \sqrt{3}\theta_l \right),
\label{k-D}
\end{equation}

\begin{equation}
\left(
\begin{array}{c}
\vec{c}_{\downarrow,j}
\\
\vec{c}_{\uparrow,j}
\end{array}
\right)
=
\sum_{\sigma=\pm}
\sum_{l=\pm}
\sum_{p=\pm}
\frac{\omega_{\sigma,l}^{pj}}
{\sqrt{J_{\sigma,l}}}
\eta^{(p)}_{\sigma,l}
\left(
\begin{array}{c}
\vec{d}^{\;\downarrow}_{\sigma,l}
\\
\vec{d}^{\;\uparrow}_{\sigma,l}
\end{array}
\right),
\label{wf-D}
\end{equation}
where $\eta$ denotes the mode amplitude, $\omega_{\pm, l}=e^{\pm i\frac{a}{2}k_l^{(\pm)}}$, 
\begin{equation}
\left[
\;^t\vec{d}_{\sigma,l}^{\;\downarrow},
\;^t\vec{d}_{\sigma,l}^{\;\uparrow}
\right]
=\left[ (\sigma l\alpha_{l}
,\;1),\;\beta_{l}
\left (\sigma l\alpha_{l} \frac{E-\varepsilon}{E+\varepsilon},\;1
\right)
\right],
\label{def-d}
\end{equation}
\begin{eqnarray}
J_{\sigma,l} &= & 
\sum_{\xi=\downarrow,\uparrow}{\rm Im}\left[w^*_{\sigma,l}
\!^t\vec{d}_{\sigma,l}^{\;\xi}h_1\vec{d}_{\sigma,l}^{\;\xi} \right]
\nonumber \\
&=& 2 |\gamma_0|v_l\sin\left(\frac{a}{2}k_l^{(\sigma)}\right),
\label{def-J}
\end{eqnarray}
\begin{equation} 
v_l=
\frac{\alpha_l}{\beta_{(-l)}}
\left(
\beta_{-}-\beta_+
\right),
\label{def-v} 
\end{equation}
\begin{equation}
\beta_{l}=\frac{2\varepsilon E - l
\sqrt{(\gamma_1^2+4\varepsilon^2)E^2-\gamma_1^2\varepsilon^2}
}{\gamma_1 (E-\varepsilon)},
\label{def-beta} 
\end{equation}
\begin{equation}
\alpha_l
=\frac{E+\varepsilon}
{\sqrt{3}|\gamma_0|\theta_l},
\label{def-alpha}
\end{equation}

\begin{equation}
\theta_l=\tau_l \frac{E}{|E|} \frac{\sqrt{E^2+\varepsilon^2
+l\sqrt{(\gamma_1^2+4\varepsilon^2)E^2-\gamma_1^2\varepsilon^2}
}}{\sqrt{3}|\gamma_0|},
\label{def-theta}
\end{equation}
where
$\tau_\pm =1$ ($\tau_\pm =\pm 1$ ) in the inner (outer) $E$ \cite{a10.,a10-2.,a10.tuika,a10-note.}.
The solid square area of Fig. \ref{fig-wavefunction}
shows the
 spatial arrangement of elements of Eq. (\ref{def-d}), where $\alpha_\downarrow = \sigma l \alpha_l$ and $\alpha_\uparrow = \alpha_\downarrow (E-\varepsilon)/(E+\varepsilon)$.
When there is a single mode such as $(\vec{c}_{\downarrow,j},\vec{c}_{\uparrow,j})
= w_{\sigma,l}^j(\vec{d}_{\sigma,l}^{\;\downarrow} ,\vec{d}_{\sigma,l}^{\;\uparrow} )$, the probability flow equals $J_{\sigma,l}$.
Here, we have chosen the definitions as conditions $v_l >0$ and $J_{\sigma,l} >0$ hold.
The probability flow of
Eq. (\ref{wf-D}) is the same as $\sum_{l,\sigma}|\eta^{(+)}_{l,\sigma}|^2-|\eta^{(-)}_{l,\sigma}|^2$, indicating that the sign $p=\pm $
denotes the propagation direction.
In the same way as the bilayer region,
the dispersion relation and wave function of each monolayer
region are represented by
\begin{equation}
\cos
\left(\frac{a}{2}k^{(\nu)}_{\;_\uparrow^\downarrow} \right) 
=\frac{-1}{2}\left(1+\nu\frac{E \pm \varepsilon}{|\gamma_0|}\right),
\label{k-mu}
\end{equation}

\begin{equation}
\vec{c}_{\downarrow,j}^{\;(0)} 
=
\sum_{\nu=\pm}
\sum_{p=\pm}
\frac{\omega_{\nu,\downarrow}^{pj}}
{\sqrt{ J_{\nu,\downarrow}}}
\eta^{(p)}_{\nu,{\rm L}}
\left(
\begin{array}{c}
\nu
\\
1
\end{array}
\right),
\label{wf-mu}
\end{equation}

\begin{equation}
\vec{c}_{\uparrow,j}^{\;(0)} 
=
\sum_{\nu=\pm}
\sum_{p=\pm}
\frac{\omega_{\nu,\uparrow}^{p(j-N)}}
{\sqrt{ J_{\nu,\uparrow}}}
\eta^{(p)}_{\nu,{\rm R}}
\left(
\begin{array}{c}
\nu
\\
1
\end{array}
\right),
\label{wf-mu-up}
\end{equation}
where $\omega_{\pm,\xi}=e^{\pm i\frac{a}{2}k_\xi^{(\pm)}}$, 
\begin{equation}
J_{\nu,\xi}=2
|\gamma_0|\sin\left(\frac{a}{2}k_\xi^{(\nu)} \right)
\end{equation}
with the layer index $\xi=\downarrow,\uparrow$.
The superscript (0) in Eqs. (\ref{wf-mu}) and (\ref{wf-mu-up})
indicates that there is no interlayer transfer integral.
Each coefficient $\eta$ of Eq. (\ref{wf-D}) 
[Eqs. (\ref{wf-mu})
and (\ref{wf-mu-up})]
corresponds to the $K_{p\sigma}$ [$K_{p\nu}$] valley .
The top and bottom of Fig. \ref{fig-wavefunction} show amplitudes of 
single layer modes in the $\downarrow\uparrow$ junction.

The scattering matrixes $S_\downarrow$ and $S_\uparrow$
are
defined by
\begin{equation}
\left(
\begin{array}{c}
\vec{\eta}^{\;(+)} \\
\vec{\eta}^{\;(-)}_{\rm L}
\end{array}
\right)
=
S_{\downarrow}
\left(
\begin{array}{c}
\vec{\eta}^{\;(-)} \\
\vec{\eta}^{\;(+)}_{\rm L}
\end{array}
\right),
\label{sl-exact}
\end{equation}
\begin{equation}
\left(
\begin{array}{c}
\Lambda^* \vec{\eta}^{\;(-)} \\
\vec{\eta}^{\;(+)}_{\rm R}
\end{array}
\right)
=
S_{\uparrow}
\left(
\begin{array}{c}
\Lambda\vec{\eta}^{\;(+)} \\
\vec{\eta}^{\;(-)}_{\rm R}
\end{array}
\right),
\label{sr-exact}
\end{equation}
where 
$\;^t\vec{\eta}^{\;(p)}=(\eta_{+,+}^{(p)},\;\eta_{-,+}^{(p)},\;\eta_{+,-}^{(p)},\;\eta_{-,-}^{(p)})
$, $\;^t\vec{\eta}^{\;(p)}_{\rm L}
=(\eta_{+,{\rm L}}^{(p)},\;\eta_{-,{\rm L}}^{(p)})
$,
$\;^t\vec{\eta}^{\;(p)}_{{\rm R}}
=(\eta_{+,{\rm R}}^{(p)},\;\eta_{-,{\rm R}}^{(p)})
$ and

\begin{equation}
\Lambda_{(\sigma,l|\sigma',l')}
=
\delta_{\sigma,\sigma'}\delta_{l,l'}e^{\sigma i\frac{a}{2}Nk_{l}^{(\sigma)}}.
\label{exact-lambda} 
\end{equation}
The boundary conditions are 
\begin{equation}
(\vec{c}_{\downarrow,0},\vec{c}_{\uparrow,0},\vec{c}_{\downarrow,1})
=(\vec{c}_{\downarrow,0}^{\;(0)},0,\vec{c}_{\downarrow,1}^{\;(0)})
\label{sl-boundary}
\end{equation}
for $S_\downarrow$ and
\begin{equation}
(\vec{c}_{\uparrow,N},\vec{c}_{\downarrow,N}, \vec{c}_{\uparrow,N-1})
=(\vec{c}_{\uparrow,N}^{ \;(0)} , 0,\vec{c}_{\uparrow,N-1}^{ \;(0)} )
\label{sr-boundary}
\end{equation}
for $S_\uparrow$. 
The exact $S_\downarrow$ and $S_\uparrow$ 
follow from the application of Eqs. (\ref{wf-D}), (\ref{wf-mu}), and (\ref{wf-mu-up}) to Eqs. (\ref{sl-boundary}) and (\ref{sr-boundary}).
Eliminating $\eta^{(\pm)}$ from Eqs. (\ref{sl-exact}) and (\ref{sr-exact}), 
we obtain \begin{equation}
t_{\uparrow\downarrow}=\;^tt_{\uparrow}\Lambda
\left({\bf 1}_4-r_{\downarrow}\Lambda r_{\uparrow}\Lambda\right)^{-1}
t_{\downarrow},
\label{sl-sr}
\end{equation}
which satisfies $\vec{\eta}^{\;(+)}_{\rm R}=t_{\uparrow\downarrow}\vec{\eta}^{\;(+)}_{\rm L}$,
where $r_{\xi}$ ($t_{\xi}$) and ${\bf 1}_n$ denote 
the $S_{\xi}$ reflection (transmission) block and the $n$
 dimensional unit matrix,   respectively.
The transmission rate $T_{\nu',\nu}(0)$ 
equals the squared absolute value of the element of $t_{\uparrow\downarrow}$.
Replacing $\uparrow$ with $\downarrow$ in Eq. (\ref{sl-sr}),
we obtain $T_{\nu',\nu}$ of the $\downarrow\downarrow$ junction.

\subsection{First approximation }
The approximation $k_\pm^{(\sigma)}
\simeq 
k_\downarrow^{(\sigma)}
\simeq \frac{4\pi}{3a}
$ 
simplifies Eq. (\ref{sl-boundary}) to
\begin{equation}
X_\downarrow
\left(
\begin{array}{c}
\vec{\eta}^{\;(+)} \\
\vec{\eta}^{\;(-)}_{\rm L}
\end{array}
\right)
=
-X_\downarrow^*
\left(
\begin{array}{c}
\vec{\eta}^{\;(-)} \\
\vec{\eta}^{\;(+)}_{\rm L}
\end{array}
\right),
\end{equation}
where 
\begin{equation}
X_\downarrow
= 
\left(
\begin{array}{ccc}
u_+^{\downarrow}
& 
u_-^{\downarrow}
& 
-u_0
\\
u_+^{\uparrow}
& 
u_-^{\uparrow}
& 
0
\\
u_+^{\downarrow}\Omega
& 
u_-^{\downarrow}\Omega
& 
-u_0\Omega^*
\end{array}
\right),
\end{equation}
\begin{equation}
u_l^{\xi}
=\frac{1}{\sqrt{v_l}}(\vec{d}_{+,l}^{\;\xi},\vec{d}_{-,l}^{\;\xi})
,\;\;
u_0
= 
\left(
\begin{array}{cc}
1 & -1 \\
1 & 1
\end{array}
\right),
\end{equation}
\begin{equation}
\Omega
= 
\left(
\begin{array}{cc}
e^{i\frac{2}{3}\pi}
& 0 \\
0 & e^{-i\frac{2}{3}\pi}
\end{array}
\right).
\end{equation}
Using the Kronecker product, we present
$S_\downarrow=-X_\downarrow^{-1}X_\downarrow^*$ 
as
\begin{equation}
S_{\downarrow}=
Y_B\;^tY_B\otimes u_B
+Y_A\;^tY_A\otimes u_A
-{\bf 1}_6,
\label{sl} 
\end{equation}
where 
\begin{equation}
Y_B
= 
\sqrt{\frac{2}{c_B}}
\left(
\begin{array}{cc}
\frac{\alpha_+}{\sqrt{v_+}}
\\
- \frac{\alpha_-}{\sqrt{v_-}}
\\
1
\end{array}
\right)
,\;
Y_A
= 
\sqrt{\frac{2}{c_A}}
\left(
\begin{array}{cc}
\frac{1}{\sqrt{v_+}}
\\
\frac{1}{\sqrt{v_-}}
\\
1
\end{array}
\right),
\end{equation}

\begin{equation}
u_{\;_A^B}
= 
\; \frac{1}{2} \left(
\begin{array}{cc}
1 & \pm 1
\\
\pm 1 & 1
\end{array}
\right),
\end{equation}
\begin{equation}
c_{\;_A^B}=
1+\frac{\alpha_+^{\pm 1} \beta_-+\alpha_-^{\pm 1} \beta_+}{\beta_--\beta_+}.
\end{equation}
We can easily confirm
that Eq. (\ref{sl}) satisfies relation $X_\downarrow S_\downarrow=
-X_\downarrow^*$. 
We also obtain an approximate formula 
\begin{equation}
S_\uparrow= V S'_\downarrow V,
\label{sr} 
\end{equation}
where
\begin{equation}
V
= \left(
\begin{array}{ccc}
\frac{\beta_+}{|\beta_+|}{\bf 1}_2 & 0 & 0
\\
0 & \frac{\beta_-}{|\beta_-|}{\bf 1}_2 & 0
\\
0 & 0 & {\bf 1}_2
\end{array}
\right),
\end{equation}
and we transform $S_\downarrow$ into $S'_\downarrow$
by replacing $(\alpha_l,\beta_l)$
with $(\alpha_l',\beta_l')=\left(\frac{E-\varepsilon}{E+\varepsilon}\alpha_l, \frac{1}{\beta_l}\right)$. 
Equations (\ref{sl}) and (\ref{sr}) satisfy
 the unitary condition $\!^tS_\downarrow^*S_\downarrow=\!^tS_\uparrow^*S_\uparrow=
{\bf 1}_6$.
The calculation of $T_{\nu',\nu}(0)$ with Eqs. (\ref{exact-lambda}),
(\ref{sl-sr}), (\ref{sl}), and (\ref{sr}) is referred to as the first approximation.

\subsection{Second approximation} 
Figure \ref{fig-band} and Eq. (\ref{k-D}) demonstrate that 
$\frac{a}{2}k_l^{(\pm)} \simeq \frac{2\pi}{3} \pm l\theta_l$.
This leads to an approximation that replaces Eq. (\ref{exact-lambda}) 
 as follows:
\begin{equation}
\Lambda_{(\sigma,s|\sigma',s')}
=
\delta_{\sigma,\sigma'}\delta_{s,s'}\exp\left[iN \left(\sigma\frac{2}{3}\pi 
+l\theta_l\right) \right].
\label{appro-lambda-2} 
\end{equation}
When $N/3$ is an integer, 
the phase $2\sigma\pi N/3$ of Eq. (\ref{appro-lambda-2}) 
has no effect. This simplifies Eq. (\ref{sl-sr}) to
\begin{equation}
t_{\uparrow\downarrow}=
\sum_{\chi=A,B} 
\;^tt_{\uparrow,\chi} \tilde{\Lambda}
\left({\bf 1}_2-r_{\downarrow,\chi}\tilde{\Lambda} r_{\uparrow,\chi}
\tilde{\Lambda}\right)^{-1}
t_{\downarrow,\chi}\otimes u_\chi,
\label{sl-sr-2}
\end{equation}
where
\begin{equation}
\tilde{\Lambda}=
\left(
\begin{array}{cc}
e^{i\theta_+ N}
& 0 \\
0 & e^{-i\theta_- N}
\end{array}
\right).
\end{equation}
See  Sect. 3.2 for the definitions  of  the $t_{\xi,\chi}$ and $r_{\xi,\chi}$ matrixes.
For example, $\;^tt_{\downarrow, A}=\frac{2}{c_A}\left( \frac{1}{\sqrt{v_+}},\frac{1}{\sqrt{v_-}}\right)$ and 
$r_{\downarrow, A}= \frac{c_A}{2}t_{\downarrow, A}\;^tt_{\downarrow, A} -{\bf 1}_2$.
The relation $u_\chi u_{\chi'} =\delta_{\chi,\chi'} u_\chi$ 
and Eq. (\ref{sl-sr-2}) bring us
the second approximation
\begin{equation}
T_{\nu',\nu}=
\left| 
\frac{G_B}{F_B} +\nu\nu'
\frac{G_A}{F_A}
\right|^2.
\label{3N-T}
\end{equation}
See Appendix B for explicit expressions
of $G$ and $F$.
Note that the second approximation (\ref{3N-T}) is valid
only when $N/3$ is an integer.

Here, we introduce
\begin{equation}
q \equiv \left(\frac{2\varepsilon}{\gamma_1} \right)^2.
\label{q-theta}
\end{equation}
When $|E|$ approaches the gap edge $\Delta$,
$|\theta_\pm|$ converges to $\theta_0$, where
\begin{equation}
\theta_0 =
\frac{\gamma_1 }{2\sqrt{3}|\gamma_0|}\sqrt{\frac{q(2+q)}{1+q}}.
\label{theta0}
\end{equation}
At the gap edge $|E|=\Delta$, 
\begin{equation}
\left(
T_{\pm,\pm}, T_{\pm,\mp}
\right)
\simeq
\frac{q}{1+q}
\left(\frac{\sin^2(\theta_0N)}{\theta_0^2N^2},\;\cos^2(\theta_0N)
\right)
\label{3N-T-2}
\end{equation}
holds in the $N$ range
\begin{equation}
\frac{\sqrt{(2+q)^3}}{(2+2q) \theta_0} \ll
N \ll 2\frac{\sqrt{2+q}}{ \theta_0}.
\label{cond-theta0}
\end{equation}
This suggests that 
\begin{equation}
g'_{\rm v}(0)
\simeq \frac{q}{1+q} 
\label{tuika-q} 
\end{equation}
when conditions (\ref{cond-theta0}),
\begin{equation}
N \simeq \pi/\theta_0,
\label{Np}
\end{equation}
and $|E| \simeq \Delta$ 
hold. The exact $g'_{\rm v}(0)$ peak
certainly appears under conditions (\ref{cond-theta0}), (\ref{Np}), and
\begin{equation}
|E| \simeq \frac{\Delta}{3}\sqrt{11+q}
\label{Ep}
\end{equation}
with the height of about $q/(1+q)$. 
Appendix B shows that Eq. (\ref{Np})
requires the condition $q<8$.
As $|E|$ increases from $\Delta$ to Eq. (\ref{Ep})
with $N$ fixed to Eq. (\ref{Np}),
$N(|\theta_+|+|\theta_-|)/2$ remains about $\pi$,  whereas
 the difference between $\pi$ and $N|\theta_{\pm}|$ 
increases from near zero to about $\pi/3$.

When $|E| \gg \sqrt{4\varepsilon^2+\gamma_1^2}$, 
Eq. (\ref{3N-T}) approximates
\begin{equation}
T_{\pm,\pm}
\simeq
\frac{(s_+ - s_-)^2}{\left|F'\right|^2 (1+q)}
,\;\;
T_{\pm,\mp}
\simeq
\frac{\gamma_1^2(s_++s_-)^2}
{4 \left|F'\right|^2 E^2},
\label{3N-T-3}
\end{equation}
where $s_\pm = \sin(|\theta_{\pm}|N)$ and
\begin{equation}
F' =\frac{-1}{1+q}\sin^2\left( 
\frac{|\theta_+|-|\theta_-|}{2} N 
\right)
-e^{i(|\theta_+|+|\theta_-|)N}.
\end{equation}
Refer to Appendix B for the derivation.
The second approximate $T_{\pm,\mp}$ formula of the $\downarrow\downarrow$ junction (not shown here explicitly) also becomes proportional to $E^{-2}$ when $|E| \gg \sqrt{4\varepsilon^2+\gamma_1^2}$.

The second approximation is closely related
to another expression of Eq. (\ref{wf-D}),
\begin{equation}
\left(
\begin{array}{c}
\vec{c}_{\downarrow,j}
\\
\vec{c}_{\uparrow,j}
\end{array}
\right)
=
\sum_{\chi =A,B}\sum_{l=\pm}\sum_{p=\pm}
\frac{e^{ilp\theta_l j}}
{\sqrt{\sqrt{3}|\gamma_0| v_l}}
\zeta^{(p)}_{\chi,l}
\vec{f}^{\;(p)}_{\chi,l,j},
\label{wf-D-AB}
\end{equation}
where
\begin{equation}
\left(
\vec{f}_{B,l,j}^{\;(\pm)},
\vec{f}_{A,l,j}^{\;(\pm)}
\right)
=
\left(
\begin{array}{cc}
0& \alpha_l
\\
1& 0
\\
0& \beta_l\alpha_l'
\\
\beta_l,& 0
\end{array}
\right)
\left(
\begin{array}{cc}
c_j& \pm is_j
\\
\pm ils_j & lc_j
\end{array}
\right),
\label{def-vec-f}
\end{equation}
$\zeta_{\;_A^B,l}^{(p)}=\eta_{+,l}^{(p)}\pm\eta_{-,l}^{(p)}$,
$c_j=\cos(2\pi j/3)$, and $s_j=\sin(2\pi j/3)$.
Equations (\ref{wf-D-AB}) and (\ref{def-vec-f}) possess two important
characteristics.
First, mode $\zeta_{\chi,l}^{(p)}$ is a Bloch
state with the unit cell length $3a/2$ and the wave number $2 pl\theta_l/a$.
Second, $\vec{f}_{\chi,l,j}^{\;(p)}$ 
is localized at the $\chi$ site when $j/3$ is an integer.
These characteristics explain
the interference between the $\vec{f}_A$ and $\vec{f}_{B}$
modes in Eq. (\ref{3N-T}).

\section{ Results}

\subsection{$\downarrow\uparrow$ junction}
\begin{figure}
\begin{center}
\includegraphics[width=0.7\linewidth]{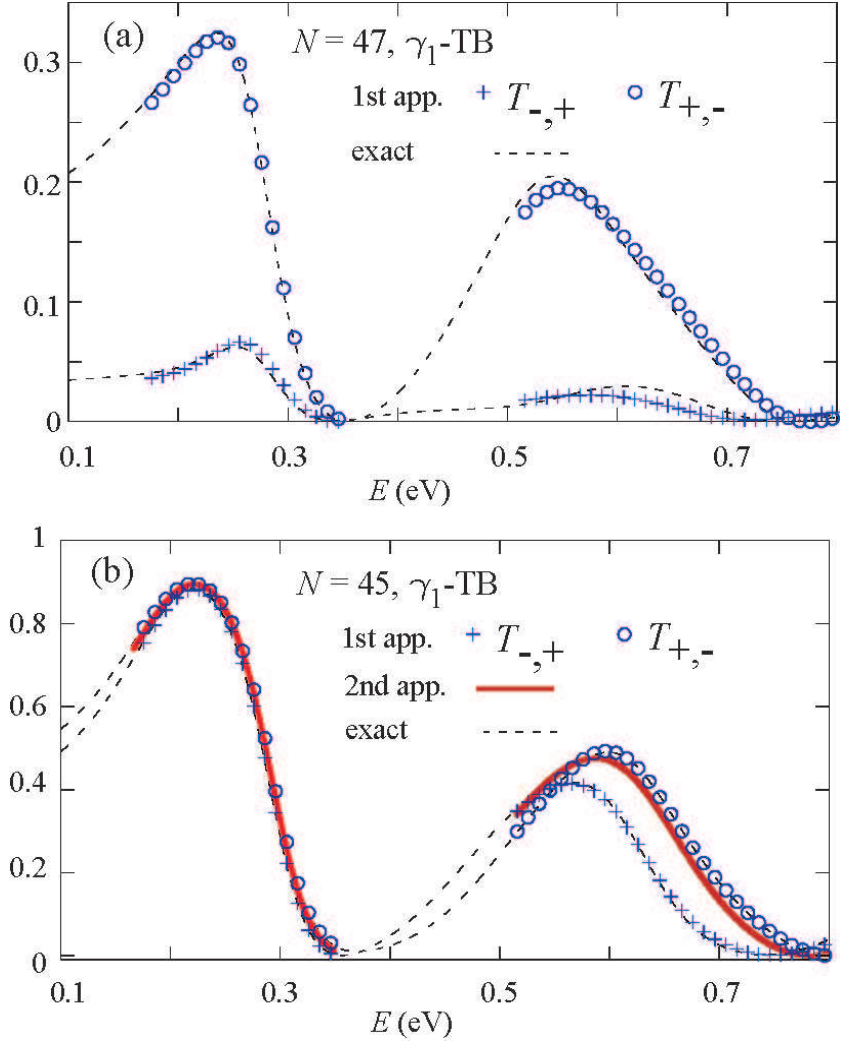}
\caption{(Color online) VCR transmission rate $T_{\pm,\mp}$ 
calculated using the $\gamma_1$-TB model in the case of $\kappa=0$, $\varepsilon=$ 0.35 eV,
(a) $N=47$, and (b) $N=45$.
}\label{fig-1stapp}
\end{center}

\end{figure}

Crosses and circles in Fig. \ref{fig-1stapp}
represent the first approximations of $T_{-,+}(0)$ and $T_{+,-}(0)$, respectively.
The longitudinal overlap lengths are $N=47$ 
in Fig. \ref{fig-1stapp}(a) and $N=45$ in Fig. \ref{fig-1stapp}(b).
The solid line in Fig. \ref{fig-1stapp}(b)
shows the second approximation
of $T_{\pm,\mp}$.
As the second approximation is irrelevant to a non-integer $N/3$,
the solid line does not appear in Fig. \ref{fig-1stapp}(a).
These approximation results reproduce well
the exact $T_{+,-}(0)$ and $T_{-,+}(0)$ displayed by
the dashed lines.
Unlike the exact calculation, the approximate formulas
are not available in the gap and pseudogap energy regions.
However, these energy regions
are unimportant because of their low transmission rate.
As Fig. \ref{fig-1stapp} shows, the difference between $T_{+,-}(0)$ and $T_{-,+}(0)$ is smaller
with an integer $N/3$ than with a non-integer $N/3$.
It follows that an integer $N/3$
is advantageous for the $g'_{\rm v}(0)$ to reach the upper limit.
In the $\gamma_1\gamma_3\gamma_4$-TB model,
Eq. (\ref{sym1}) is exact, whereas  Eq. (\ref{sym2}) is approximate; thus,
 the relation $\tilde{g}_{\rm v}(E,\varepsilon) =
\tilde{g}_{\rm v}(-E,\varepsilon) $ does not exactly hold.
The numerical results show that the $\tilde{g}_{\rm v}$
peak is slightly higher in the negative $E$
region than in the positive $E$ region.
As high $\tilde{g}_{\rm v}$ is our concern, 
 the figures in this paper, except for Figs. \ref{fig-band} and \ref{fig-1stapp}, display data on an integer $N/3$ and negative $E$.

\begin{figure}
\begin{center}
\includegraphics[width=0.7\linewidth]{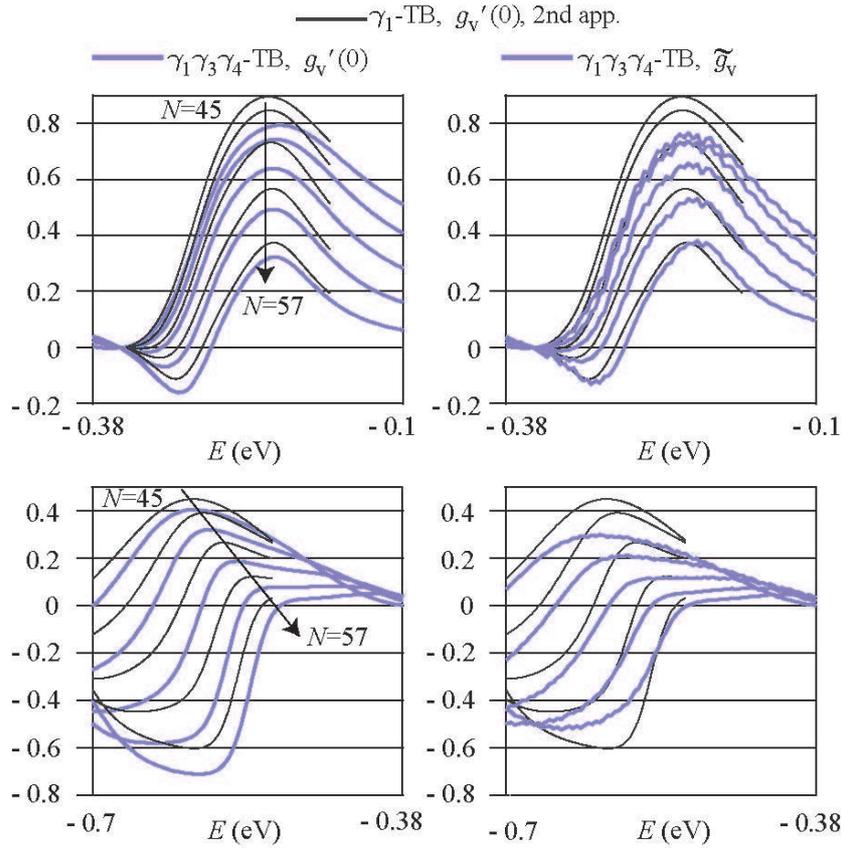}
\caption{
(Color online) VCR data of the $\downarrow\uparrow$ junction
in the case of $\varepsilon=0.35$ eV, $N_y =1000$,
and $N=45,48,51,54,$ and 57.
Blue lines in the left and right panels represent 
the $\gamma_1\gamma_3\gamma_4$-TB data of
$g'_{\rm v}(0)$ 
and $\tilde{g}_{\rm v}$, respectively.
Black lines represent the second approximation of $g'_{\rm v}(0)$.
The top and bottom panels 
mainly correspond to the inner and outer $E$ regions, respectively.
}\label{fig-LDRe70}
\end{center}
\end{figure}
\begin{figure}
\begin{center}
\includegraphics[width=0.7\linewidth]{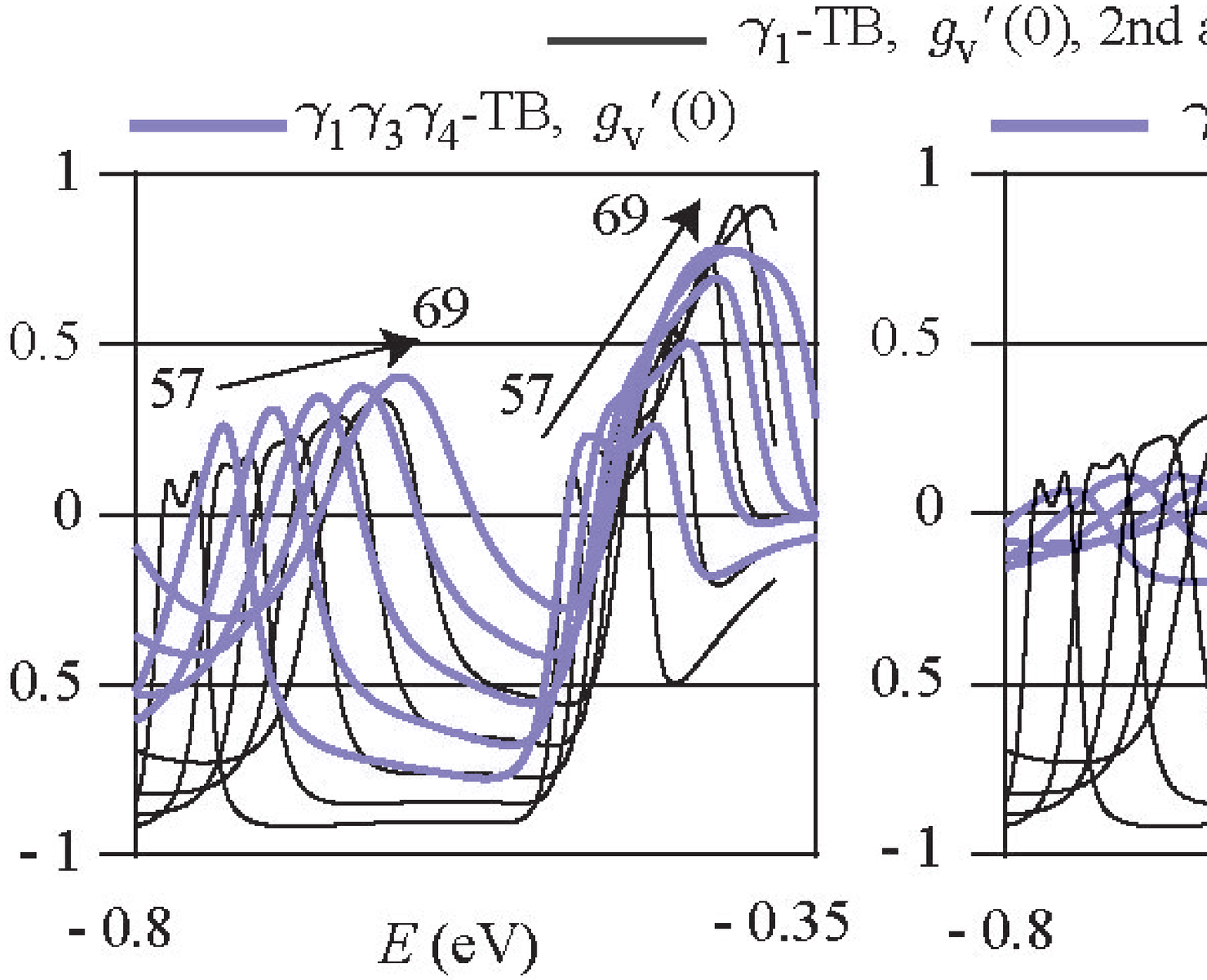}
\caption{(Color online) Data of $g'_{\rm v}(0)$ 
and $\tilde{g}_{\rm v}$ in the case of $\varepsilon=0$, $N_y =1000$,
 and $N=57,60,63,66,$ and 69.
Blue lines in the left and right panels represent 
the $\gamma_1\gamma_3\gamma_4$-TB data of
$g'_{\rm v}(0)$ 
and $\tilde{g}_{\rm v}$, respectively.
Black lines represent the second approximation of $g'_{\rm v}(0)$.
}\label{fig-LDRe0}
\end{center}

\end{figure}

\begin{figure}
\begin{center}
\includegraphics[width=0.7\linewidth]{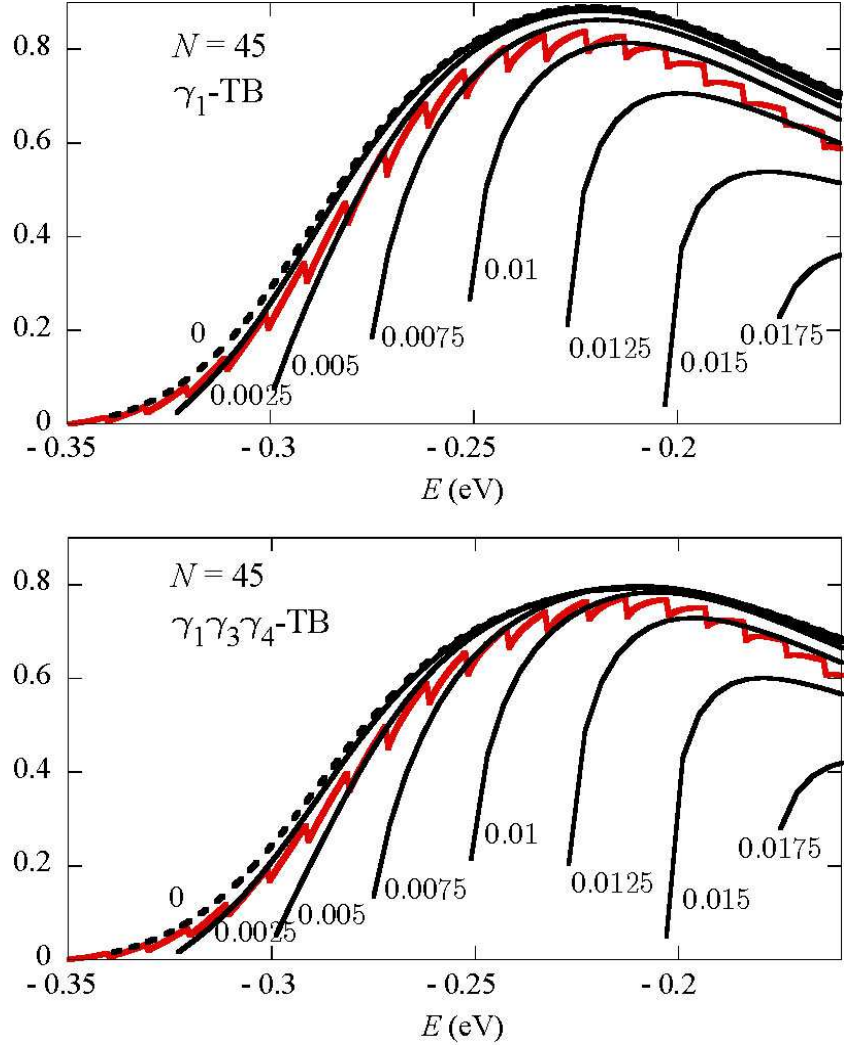}
\caption{(Color online) $\tilde{g}_{\rm v}$ (red lines) 
and $g'_{\rm v}(\kappa)$ (black lines) in the case of $\varepsilon= 0.35$ eV, $N=45$,
 and 
 $N_y =1000$, calculated using the $\gamma_1$-TB (top panel) and $\gamma_1\gamma_3\gamma_4$-TB models (bottom panel).
The numerical values attached to each black line represent
$\kappa/\pi$.
}\label{fig-gke70}
\end{center}

\end{figure}

Figures \ref{fig-LDRe70} 
and \ref{fig-LDRe0}
show the data on $\varepsilon=0.35$ eV
and $\varepsilon=0$, respectively.
$N$ increases from 45 to 57 (from 57 to 69)
in Fig. \ref{fig-LDRe70} (Fig. \ref{fig-LDRe0}),
where $N$ is limited to a multiple of three.
In Fig. \ref{fig-LDRe70}, $E$ changes from $-0.38$ eV to $-0.1$ eV
(from $-0.7$ eV to $-0.38$ eV ) in the top (bottom) panels.
The $E$ ranges partially overlap the gap $|E| < 0.166$ eV
and pseudogap 0.35 eV $ < |E| < 0.514$ eV
in Fig. \ref{fig-LDRe70}.
Figure \ref{fig-LDRe0}
includes a part of the pseudogap $|E| <$ 0.377 eV.
Blue lines represent the $\gamma_1\gamma_3\gamma_4$-TB data, indicating $g'_{\rm v}(0)$ in the left panels
and $\tilde{g}_{\rm v}$ in the right panels.
The left panels are identical to the right panels
in the black lines representing the second approximation of $g'_{\rm v}(0)$.
Each pair of identical black lines helps compare
the left and right panels in the blue lines.
In the gap and pseudogap,  there are no black lines,
but the blue lines confirm the suppression of $g'_{\rm v}(0)$ and
$\tilde{g}_{\rm v}$.
The second approximation excellently reproduces
the $T_{\nu,\nu'}(0)$ of the exact $\gamma_1$-TB calculation, 
as Fig. \ref{fig-1stapp} shows.
Thus, $\gamma_2$ and $\gamma_3$ are the leading causes 
of the slight difference between the black and blue lines in 
the left panels and have only minor effects on $g'_{\rm v}(0)$.
On the other hand, $g'_{\rm v}(\kappa)$ with nonzero $\kappa$ 
causes differences between 
the left and right panels in blue lines.
Figure \ref{fig-gke70}
displays the decomposition of  $\tilde{g}_{\rm v}$ 
into $g'_{\rm v}(\kappa)$
and elucidates the $\kappa$ effect 
for the highest $\tilde{g}_{\rm v}$ line of Fig. \ref{fig-LDRe70}
(top panel, $N=45$).
The top and bottom panels correspond to the $\gamma_1$-TB and 
$\gamma_1\gamma_3\gamma_4$-TB models, respectively.
The solid black lines represent
$g'_{\rm v}(\kappa)$ 
for seven $\kappa'$s, $\kappa/\pi= 0.0025m$ ($m=1,2,\cdots,7)$.
The black dashed lines represent the exact $g'_{\rm v}(0)$
and are 
essentially the same as the second approximation of $g'_{\rm v}(0)$.
The eight black lines contribute to $\tilde{g}_{\rm v}$ 
when $N_y=400$ and $ -0.35$ eV $< E < -0.16$ eV.
As has been defined by Eqs. (\ref{g-gvr}) and (\ref{ave-gv}), 
$\tilde{g}_{\rm v}$ equals the sum of $g'(\kappa)$ divided by $2M+1$,
where the channel number $2M+1$ changes with $E$ according to Eq. (\ref{M-condition}).
For example, $(E ,2M+1)=(-0.34$ eV, 1) and $(-0.26$ eV, 7)
when $N_y=400$. 
The red lines display 
$\tilde{g}_{\rm v}$
computed for the width $N_y=1000$, where the small notches reflect the finite $N_y$. 
As $N_y$ increases, the $\tilde{g}_{\rm v}$ line becomes
smooth and independent of $N_y$.
The monolayer energy gap $||E|-\varepsilon| < |\gamma_0 \sin \kappa|$,
 which originates from Eq. (\ref{monolayer-gap}),
widens as $|\kappa|$ increases,
 and thus, only small $|\kappa|$'s contribute to
the red lines in Fig. \ref{fig-gke70}.
In this small range of $\kappa$, 
$\tilde{g}_{\rm v}$ is close to 
$g'_{\rm v}(0)$ as $g'_{\rm v}(\kappa)$ is continuous
with respect to $\kappa$. 
This is the reason why the blue lines of the left panels
are similar to those of the right panels in Fig. \ref{fig-LDRe70}.
Figure \ref{fig-gke70} also proves that 
the $\gamma_1$-TB and 
$\gamma_1\gamma_3\gamma_4$-TB models
produce essentially the same $\tilde{g}_{\rm v}$
except that $\gamma_3$ and $\gamma_4$ lower the peak slightly.
In Ref. \cite{tamura-2019} , this smallness of $\gamma_3,\gamma_4$ effects
 is discussed.
These results indicate that
the second approximation $g'_{\rm v}(0)$ is very close to
$\tilde{g}_{\rm v}$.
In the case of $\varepsilon=0$, however, this is not true
as discussed below.

\begin{figure}
\begin{center}
\includegraphics[width=0.7\linewidth]{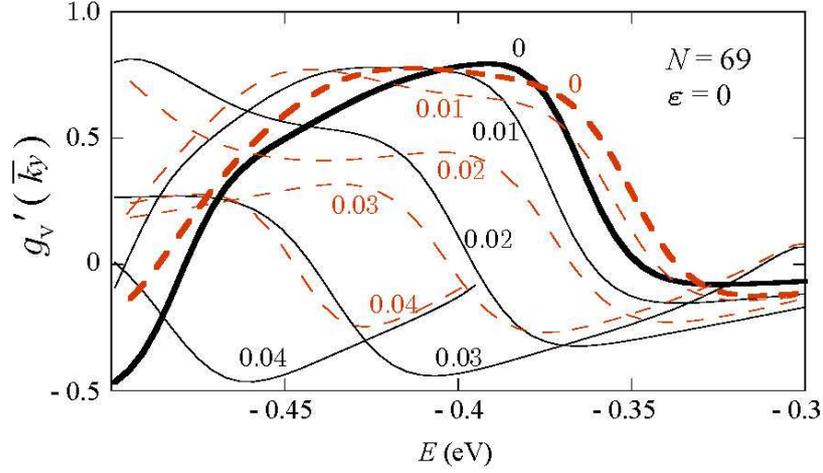}
\caption{(Color online) $g'_{\rm v}(\kappa)$ in the case of $\varepsilon=0$, and $N=69$.
The numerical values attached to each line represent
$\kappa/\pi$.
The solid and dashed lines 
represent values  calculated using the $\gamma_1$-TB and 
$\gamma_1\gamma_3\gamma_4$-TB models, respectively.
Thick (thin) lines correspond to zero (nonzero) $\kappa$. 
}\label{fig-gke0}
\end{center}

\end{figure}

Figure \ref{fig-gke0}
displays
$g'_{\rm v}(\kappa)$ of zero $\varepsilon$ 
for five $\kappa$'s, $\kappa/\pi= 0, 0.01 ,\cdots,0.04$
in case $\varepsilon=0$ and $N=69$, 
corresponding
to the highest $\tilde{g}_{\rm v}$ peak in Fig. \ref{fig-LDRe0}.
The solid and dashed lines are calculated using the $\gamma_1$-TB and 
$\gamma_1\gamma_3\gamma_4$-TB models, respectively.
Thick (thin) lines correspond to zero (nonzero) $\kappa$. 
In the pseudogap $|E| < \gamma_1$, 
$g'_{\rm v}(\kappa)$ is suppressed.
The peak height of 
$g'_{\rm v}(0)$ is comparable to that in Fig. \ref{fig-gke70}.
Compared with Fig. \ref{fig-gke70}, however, a wider range of $\kappa$ contributes to $\tilde{g}_{\rm v}$ in Fig. \ref{fig-LDRe0}.
When $\tilde{g}_{\rm v}$ reaches the maximum,
the maximum  effective $|\kappa|/\pi$ equals 0.013
in Fig. \ref{fig-LDRe70} and 0.045 in Fig. \ref{fig-LDRe0}.
As $\kappa$ increases, 
the $g'_{\rm v}(\kappa)$ of Fig. \ref{fig-gke0}
decreases to negative values
and lowers the maximum $\tilde{g}_{\rm v}$ peak height 
shown in Fig. \ref{fig-LDRe0} compared with that shown in  Fig. \ref{fig-LDRe70}.

Arrows indicate
the shift of the peaks with $N$
in Figs. \ref{fig-LDRe70}
and \ref{fig-LDRe0}.
In the outer $E$,
the peak energy $E$ approaches
zero as $N$ increases.
This shift comes from the phases $|\theta_\pm|N$ of Eq. (\ref{3N-T-3}),
where $|\theta_\pm|$ increases with $|E|$ in the outer $E$.
In the inner $E$, however, only the peak height changes
with an almost constant peak energy 
(top panels in Fig. \ref{fig-LDRe70}).
This contrast suggests
the difference between the outer and inner $E$ regions
in the VCR origin.
As shown in Figs. \ref{fig-intro} and \ref{fig-wavefunction} ,
$I_\xi(\nu) \equiv |(\nu,1) \vec{d}^{\;\xi}_{\sigma,l}|^2/|\vec{d}^{\;\xi}_{\sigma,l}|^2=(1+\nu\alpha_\xi)^2/(1+\alpha_\xi^2)$
represents the monolayer $\xi$--bilayer matching, where $\xi =\downarrow,\uparrow$, $\alpha_\downarrow =\sigma l \alpha_l$, and $\alpha_\uparrow= \alpha_\downarrow(E-\varepsilon)/(E+\varepsilon)$. 
The product of monolayer--bilayer matchings $I_\downarrow(\pm)I_\uparrow(\mp)$ corresponds to
the VCR and comes near the upper limit when (i) $\alpha_\downarrow \simeq -\alpha_\uparrow \simeq \pm 1$. Since 
$ \alpha_\uparrow/\alpha_\downarrow =(E-\varepsilon)/(E+\varepsilon)$
, condition (i) requires $E\simeq 0$.
The gap region $|E| < \Delta$ suppresses the transport near the zero energy, and thus
the optimized $E$ comes at the gap edge $\pm \Delta$, followed by $\alpha_\downarrow=\sigma l ( 1 \pm \sqrt{1+q})/\sqrt{2+q}$. 
When $q \gg 1$, this $\alpha_\downarrow$ satisfies condition (i). 
The condition $E \simeq \Delta$ is independent of $N$
and corresponds to the constant peak energy in the top panels of Fig. \ref{fig-LDRe70}.
For a large interlayer transmission rate,
the wave function must be extended between the two layers,
and thus the interlayer probability ratio
$\tilde{\beta}_l \equiv
|\vec{d}_{\sigma,l}^{\;\uparrow}|/
\vec{d}_{\sigma,l}^{\;\downarrow}| $
must be close to one.
We define the interlayer matching $I_{\downarrow\uparrow} \equiv 4/(\tilde{\beta}_l+\tilde{\beta}_l^{-1})^2$, 
since this $I_{\downarrow\uparrow}$ increases as $\tilde{\beta}_l$ approaches one.
Interestingly, 
$I_{\downarrow\uparrow}$ is close to Eq. (\ref{tuika-q}) when $|E|\simeq \Delta \ll \varepsilon$ and $\tilde{\beta}_l \simeq |\beta_l|$,
suggesting the contributions of $\beta_l$ to the VCR.
When $|E| \gg \Delta'$, on the other hand,
the monolayer--bilayer matching product $I_\downarrow(\pm) I_\uparrow(\mp) $ becomes inversely proportional to $E^2$, implying the physical origin of the factor $E^{-2}$ in Eq. (\ref{3N-T-3}).

\begin{figure}
\begin{center}
\includegraphics[width=0.7\linewidth]{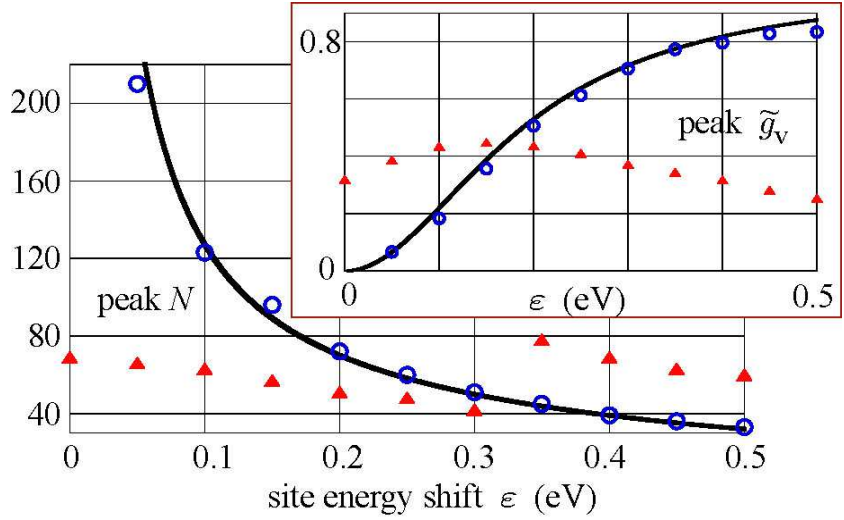}
\caption{(Color online) Height and $N$ of the highest $\tilde{g}_{\rm v}$ peak 
of the $\downarrow\uparrow$ junction calculated using the  $\gamma_1\gamma_3\gamma_4$-TB model
in the range $|E| < 1.2$ eV and $N \leq 282 $,
for eleven $\varepsilon$'s, $\varepsilon=0.05m$ eV $(m=0,1,\cdots, 10)$.
Circles and triangles correspond to inner and outer $E$ regions, 
respectively. The transverse width $3N_ya_{\rm c}$ is 3000$a_{\rm c}$.
The solid lines represent $\tilde{g}_{\rm v}=q/(1+q)$ and Eq. (\ref{Np}).
}\label{fig-NpT}
\end{center}

\end{figure}

\begin{figure}
\begin{center}
\includegraphics[width=0.7\linewidth]{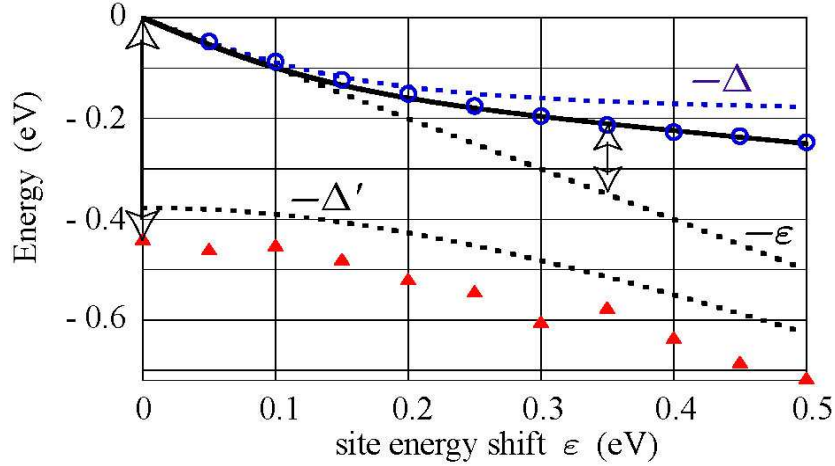}
\caption{
(Color online) Energies corresponding to the data in Fig. \ref{fig-NpT}.
The solid line represents values obtained using Eq. (\ref{Ep}).
The dotted lines show the gap and pseudogap edges.
The vertical arrows 
represent $||E|-\varepsilon|$ for the triangle of $\varepsilon=0$ and
the circle of $\varepsilon=$ 0.35 eV.
}\label{fig-Ep}
\end{center}
\end{figure}

Circles (triangles) in Figs. \ref{fig-NpT} and \ref{fig-Ep} 
 represent the highest $\tilde{g}_{\rm v}$ peak data in the inner (outer) $E$ region. These data are calculated using $\gamma_1\gamma_3\gamma_4$-TB model
in the ranges $|E| < 1.2$ eV and $N \leq 282 $,
for eleven $\varepsilon$'s, $\varepsilon=0.05m$ eV $(m=0,1,\cdots, 10)$.
Figure \ref{fig-NpT} 
shows the peak height and peak $N$. Figure \ref{fig-Ep}
shows the peak $E$.
Solid lines represent values obtained using Eqs. (\ref{tuika-q}), (\ref{Np}), 
 and (\ref{Ep}),  which accurately coincide with the circles.
The dotted lines in Fig. \ref{fig-Ep} are the gap and pseudogap
edge energies.
Figures \ref{fig-NpT} and \ref{fig-Ep} include
the inner $E$ peak $(E,N,\tilde{g}_{\rm v})
=(-0.21$ eV, 45, 0.77)
 in Fig. \ref{fig-LDRe70} 
and the outer $E$ peak $(E,N,\tilde{g}_{\rm v})=(-0.44$ eV, 69, 0.32)
 in Fig. \ref{fig-LDRe0}.
Concerning the absolute and relative VCR, $(g_{\rm v}, g_{\rm v}/g) = (0.042, 0.47)$
 for the former and (0.057, 0.24) for the latter.
Notably, the former $g_{\rm v}/g$ reaches near
the upper limit 0.5
corresponding to the perfect VCR ($T_{+,-}=T_{-,+}=1$).
Compared with the zero $\varepsilon$ case, the absolute VCR slightly decreases, but the relative VCR significantly improves.
The two vertical arrows in Fig. \ref{fig-Ep}
represent $||E|-\varepsilon|$ at the above-mentioned two peaks,
 whereas Eq. (\ref{M-condition}) indicates 
the effective $\kappa$ range $|\kappa|< ||E|-\varepsilon|/|\gamma_0$.
These arrows illustrate that the subband effects
are more significant in Fig. \ref{fig-LDRe0} 
than in Fig. \ref{fig-LDRe70}.
In the top panel of Fig. \ref{fig-NpT},
the outer $E$ peak first increases
with $\varepsilon$ but declines when $\varepsilon$ exceeds 0.15 eV.
The initial climb comes from the shrinkage of the effective
$\kappa$ range. 
As the peak $E$ becomes distant from zero, the factor $\gamma_1^2/E^2$ of
Eq. (\ref{3N-T-3}) causes the later decline.
In contrast, the inner $E$ peak continues to grow with $\varepsilon$ and 
rise above the outer peak. 
The perturbative formula in Ref. \cite{tamura-2021} is consistent with 
Eq. (\ref{3N-T-3}), but irrelevant to the inner $E$ peak \cite{note}.

\begin{figure}
\begin{center}
\includegraphics[width=0.7\linewidth]{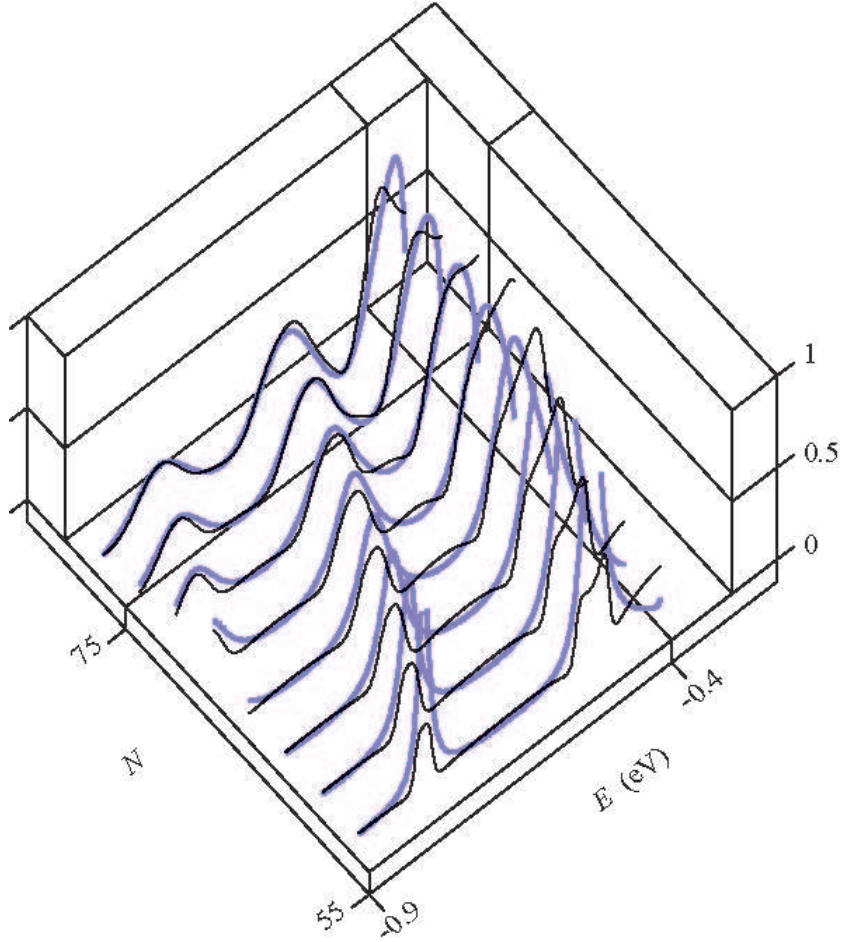}
\caption{(Color online) VCR transmission rate $T_{\pm,\mp}$ in the case of $\varepsilon=0$.
Black and blue lines represent values calculated using Eqs. (\ref{3N-T}) and (\ref{3N-T-3}), respectively.
$N$ is limited to a multiple of three.
}\label{fig-outerE}
\end{center}
\end{figure}

Equation (\ref{3N-T-3}) derives its validity from condition 
$|E| \gg \gamma_1\sqrt{1+q}$ in Appendix B, whereas Fig. \ref{fig-Ep} indicates that the outer $E$ peak
appears near the pseudogap edge, $\pm \gamma_1\sqrt{1+\frac{q}{4}}$.
This might weaken the effectiveness of 
 Eq. (\ref{3N-T-3}) discussed above.
Figure \ref{fig-outerE} dissolves this uncertainty, where 
 the black and blue lines represent the VCR transmission rate $T_{\pm,\mp}$ 
of Eqs. (\ref{3N-T}) and (\ref{3N-T-3}), respectively, in the case of $\varepsilon=0$.
Surprisingly, the blue lines coincide well with the black lines 
even when $|E|$ is close to $\gamma_1$. 
The blue lines overestimate the peak heights
for $N$ between 57 and 63, but coincide well with the black lines 
for other $N$ values. 
The decay of $T_{\pm,\mp}$ with $|E|$ certainly appears in the black lines.
Overall, Eq. (\ref{3N-T-3}) reproduces the peak position $(E,N)$
of Eq. (\ref{3N-T}).
These results mean that
 the condition $|E| \gg \gamma_1\sqrt{1+q}$ is not necessary 
but sufficient for the effectiveness of Eq. (\ref{3N-T-3}).
Although we have not obtained the necessary and sufficient
condition yet, the $E^{-2}$ factor that comes
from the monolayer--bilayer matching certainly causes the decline of the outer $E$ peak.

The $\tilde{g}_{\rm v}$
oscillates periodically as a function
of $N$, and the considered $N$ range in Fig. \ref{fig-NpT}
$(N \leq 282)$ is sufficient to include
the global maximum of $\tilde{g}_{\rm v}$.
The height of the $m$'th $g'_{\rm v}(0)$ peak at $N\simeq m\pi/\theta_0$ 
in the inner $E$ is almost constant irrespective of $m$ in the second approximation,
 but gradually decreases with $m$ in the exact $\gamma_1$-TB calculation.
This difference comes from the deviation of the average wave number $\frac{1}{2}(k^{(+)}_l+k^{(-)}_l)$ from $\frac{4\pi}{3a}$.
The second approximation neglects this deviation.
In the exact $\gamma_1$-TB calculation,
 the deviation of the phase $(k^{(+)}_l+k^{(-)}_l)Na/4$ from $\frac{2}{3}\pi N$ 
increases with $N$, followed by the decay of the peak height.
 $\gamma_3$ and $\gamma_4$ also induce the phase shift.
Additionally, the peak period in $g'_{\rm v}(\kappa)$
deviates from $\pi/\theta_0$ as $|\kappa|$ increases.
These effects cause the decay of the 
$\tilde{g}_{\rm v}$ peak with $N$ in the inner $E$.

\begin{figure}
\begin{center}
\includegraphics[width=0.7\linewidth]{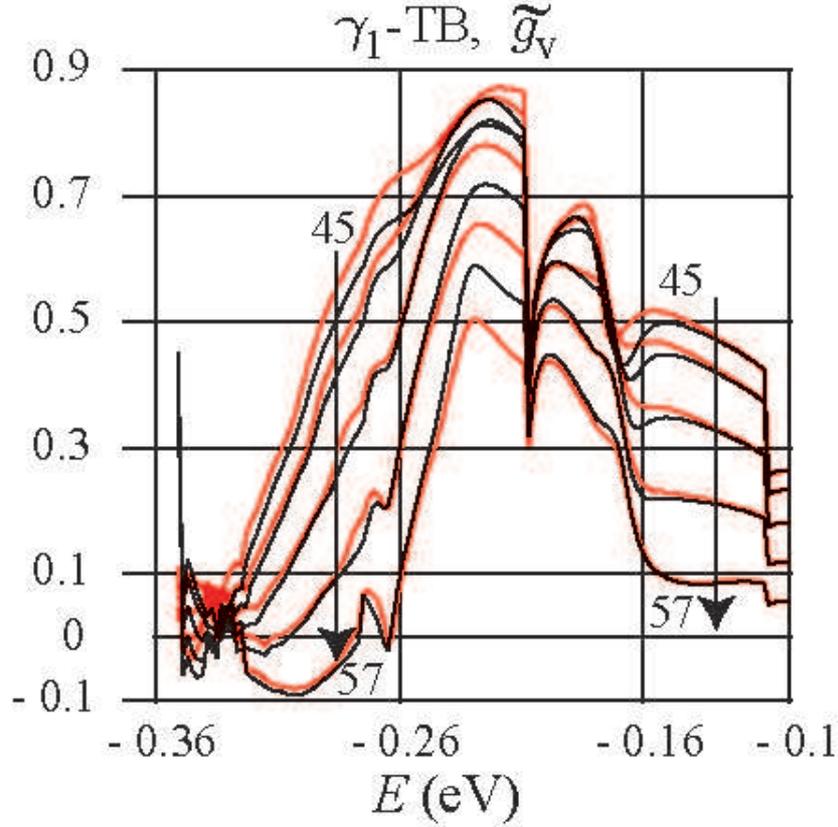}
\caption{(Color online) Exact $\tilde{g}_{\rm v}$ 
of partially overlapped (50,50) zigzag graphene ribbons
(ZGR) calculated using $\gamma_1$-TB
in the case of $N=45,48,51,54,$ and 57, and $\varepsilon =0.35$ eV.
Red and black lines
correspond to the perfect armchair edges
and the armchair edges with six addimers,
respectively. See the main text and Fig. \ref{fig-intro-xy}
 for the addimers and ZGR.
}\label{fig-defect}

\end{center}

\end{figure}

In the calculation presented so far,
we assume the perfect armchair
termination with no zigzag edge.
To see the validity of this assumption,
we replace the periodic condition with
the open boundary condition at positions $\frac{y}{a_{\rm c}}=2.5, 151.5$ 
(7.5, 156.5) of  the  $\uparrow$ layer ($\downarrow$ layer)
in the $\downarrow\uparrow$ junction. 
This junction is equivalent to partially overlapped (50,50) zigzag graphene ribbons (ZGRs).
We also consider addimers bonded with the armchair edges
as in Fig. \ref{fig-intro-xy}.
The positions of the addimers vary randomly depending
on the samples, and we consider
six addimers at $(\frac{x}{a}, \frac{y}{a_{\rm c}})=(0,32)$,(0,92),(0,122),
$(N/2,46)$, $(N/2,67)$,  and $(N/2,97)$ as a case 
of low coverage.
Figure \ref{fig-defect} corresponds to the top panels of Fig. \ref{fig-LDRe70},
where red (black) lines
indicate the exact $\gamma_1$-TB calculation
of $\tilde{g}_{\rm v}$ of the ZGR junction with no addimer
(with the six addimers).
The $\tilde{g}_{\rm v}$ curve is almost
the same as that in Fig. \ref{fig-LDRe70}
and the nearly perfect VCR survives the imperfections.
The sharp dips at $E=-0.21$ eV and $-0.11$ eV
correspond to the subband edge energies.
Additionally, a nearly flat band
of zigzag edge increases  $\tilde{g}_{\rm v}$ near
the Dirac points of single layers, $|E| \simeq \varepsilon$. 
As the ZGR width increases, however, these effects
weaken and the $\tilde{g}_{\rm v}$ curve in
 Fig. \ref{fig-LDRe70} recovers.

\subsection{ $\downarrow\downarrow$ junction}
\begin{figure}
\begin{center}
\includegraphics[width=0.7\linewidth]{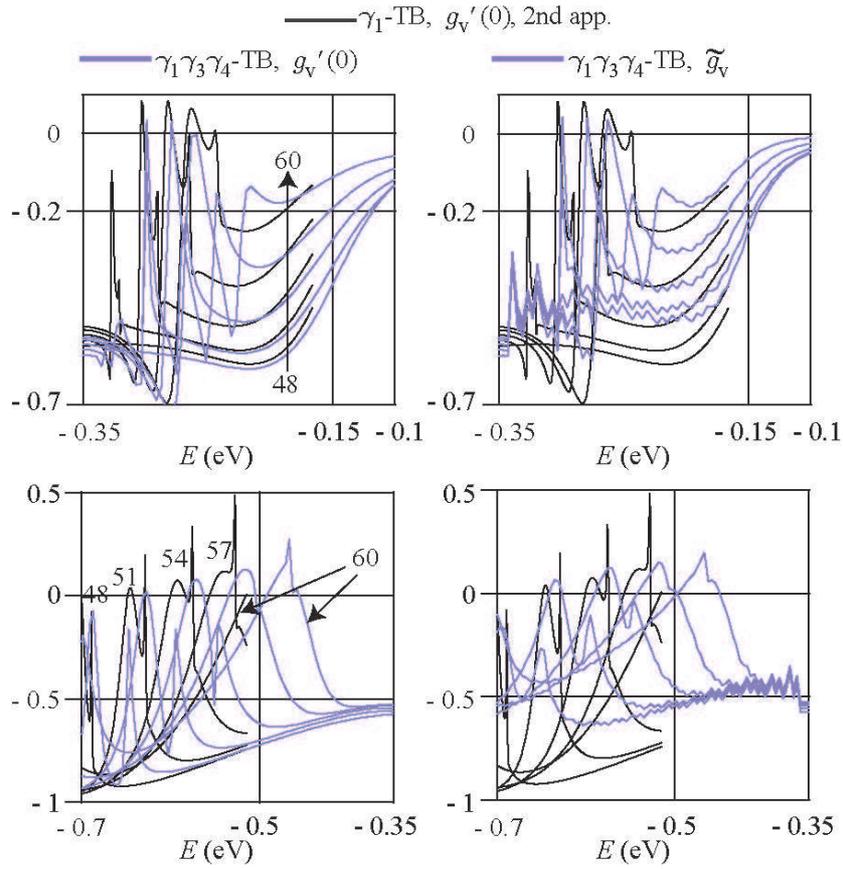}
\caption{
(Color online) VCR data of $\downarrow\downarrow$ junction
in the case of $\varepsilon=0.35$ eV, $N_y =1000$,
and $N=48,51,54,57,60$
with almost the same format as in Fig. \ref{fig-LDRe70}.
}\label{fig-LDLe70}
\end{center}
\end{figure}

\begin{figure}
\begin{center}
\includegraphics[width=0.8\linewidth]{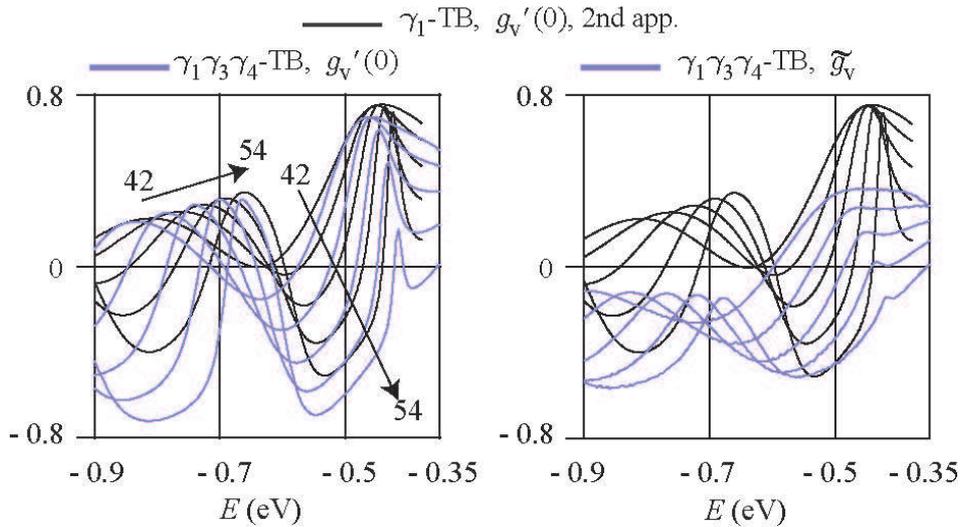}
\caption{
(Color online) VCR data of  $\downarrow\downarrow$ junction
in the case of $\varepsilon=0$, $N_y =1000$,
and $N=42,45,48,51,$ and 54
with almost the same format as Fig. \ref{fig-LDRe0}.
}\label{fig-LDLe0}
\end{center}

\end{figure}

As Eq. (\ref{M-condition}) shows, the effective $\kappa$ range $|\kappa| < |E+\varepsilon|/|\gamma_0|$
is wider for positive $E$ than for negative $E$.
It follows that the $\tilde{g}_{\rm v}$ peaks in the negative $E$ region are higher than those in the positive $E$ region.
In the case of the $\downarrow\uparrow$ junction,
an integer, $N/3$,
is a necessary condition for a near perfect $\tilde{g}_{\rm v}$
because
 the relation $T_{+,-} \simeq T_{-,+}$ 
holds under this condition.
Contrarily, the mirror symmetry of
the $\downarrow\downarrow$ junction
guarantees the relation $T_{+,-} = T_{-,+}$, irrespective of $N$.
Nevertheless, the $\tilde{g}_{\rm v}$ peak
tends to be higher
with an integer $N/3$ than a non-integer $N/3$.
We speculate that constructive interference
between the $\vec{f}_A$ and $\vec{f}_{B}$
modes in Eq. (\ref{wf-D-AB})
increases $\tilde{g}_{\rm v}$.
As our interest lies in high $\tilde{g}_{\rm v}$,
we present data on the negative $E$ and $N$
being a multiple of three.
Figures \ref{fig-LDLe70} 
and \ref{fig-LDLe0} represent the VCR data in
the cases of $\varepsilon= 0.35$ eV and $\varepsilon= 0$, respectively.
The black lines in Figs. \ref{fig-LDLe70} and \ref{fig-LDLe0} display
the second approximation of $g'_{\rm v}(0)$.
The blue lines show the exact data of $g'_{\rm v}(0)$ ($\tilde{g}_{\rm v}$) 
 calculated using the $\gamma_1\gamma_3\gamma_4$-TB model
in the left (right) panel.
Overall, Figs. \ref{fig-LDLe70} and \ref{fig-LDLe0} 
have the same format as that in Figs. \ref{fig-LDRe70} and 
\ref{fig-LDRe0}.
The good agreement between the black and blue lines in the left panels
indicates the effectiveness of the second approximation of $g'_{\rm v}(0)$ and 
 the insignificantly small effects of $\gamma_3$ and $\gamma_4$.
The $(E, N)$ ranges in Figs. \ref{fig-LDLe70} and \ref{fig-LDLe0} include or
lie around
the highest $\tilde{g}_{\rm v}$ peaks searched 
in the ranges $|E| < 1.2$ eV and $N \leq 282 $;
$(E,N, \tilde{g}_{\rm v})=(-0.47$ eV$, 60, 0.20)$ in the case of
$\varepsilon=$ 0.35 eV 
and $(-0.51$ eV, $36, 0.45)$ in the case of $\varepsilon=0$.
In Fig. \ref{fig-LDLe0}, the blue line peaks reach about 0.7 in the left panel but remain less than 0.45 in the right panel.
This reduction in $\tilde{g}_{\rm v}$
comes from the $\kappa$ effect in the same way
as Fig. \ref{fig-LDRe0}.
The monolayer--bilayer matching product 
of the $\downarrow\downarrow$ junction is defined as $I_\downarrow(\pm)I_\downarrow(\mp)$
in the same way as $I_\downarrow(\pm)I_\uparrow(\mp)$ of
the $\downarrow\uparrow$ junction.
This product $I_\downarrow(\pm)I_\downarrow(\mp)$ 
becomes also proportional to $E^{-2}$ and 
explains the similarity between Figs. \ref{fig-LDRe0} and \ref{fig-LDLe0}.
The outer $E$ peak comes near the pseudogap edge $|E|=\Delta'$
and decreases its height with $\varepsilon$
in the same way as
the outer $E$ peak of the $\downarrow\uparrow$ junction.
When $\varepsilon=0$, the outer $E$ peak of the $\downarrow\downarrow$ junction is slightly higher than that of the $\downarrow\uparrow$ junction.
When $\varepsilon $ exceeds 0.2 eV, however, the inner $E$ peak of
the $\downarrow\uparrow$ junction is dominant over the
$\downarrow\downarrow$ junction peak.

The second approximation is effective as shown in both  Figs. \ref{fig-LDLe70}
and \ref{fig-LDRe70},
where the monolayer energy gap suppresses the $\kappa$ effect.
In the inner $E$ region of $\varepsilon=$ 0.35 eV, 
$\tilde{g}_{\rm v}$ is typically
negative and about 0.1 at the most.
This reduction in $\tilde{g}_{\rm v}$ is reasonable.
As $\varepsilon$ increases, interlayer resonance
 degrades, and the $\downarrow$ layer works as
an isolate-perfect layer with no intervalley scattering.
In the case of the $\downarrow\uparrow$ junction, however, the electron must flow between the $\downarrow$ and $\uparrow$ layers, and thus,
 the $\downarrow$ layer cannot be 
isolated by increasing $\varepsilon$.
The fine notches from the finite $N_y$ effect are more visible in Fig. \ref{fig-LDLe70} than in Fig. \ref{fig-LDRe70}. 
The $g'(\kappa)$
of the $\downarrow\downarrow$ junction
suddenly changes from a considerably negative
value to zero when $|\kappa|$ exceeds 
$|E+\varepsilon|/|\gamma_0|$,
because of the $\downarrow$ layer energy gap $|E+\varepsilon| < |\gamma_0\sin\kappa|$.
This sudden change brings remarkable notches to Fig. \ref{fig-LDLe70}.
On the other hand,
$g'(\kappa)$ 
of the $\downarrow\uparrow$ junction 
with the energy $E \simeq -\varepsilon$
is close to zero, irrespective of $\kappa$.
Thus the finite $N_y$ effect is inconspicuous in Fig. \ref{fig-LDRe70}.

\section{Detection in Experiments}
The SG structure was experimentally observed in a Li-intercalated graphene\cite{(13)}. 
However, the VCR occurs in the G--SG--G double junctions, not in the SG alone \cite{graphene-precession,graphene-precession2,graphene-precession3}. 
The G--SG boundary control remains undeveloped, whereas the established G layer alignment technique is advantageous to the po-G measurement \cite{(12)}.

\begin{figure}
\begin{center}
\includegraphics[width=0.7\linewidth]{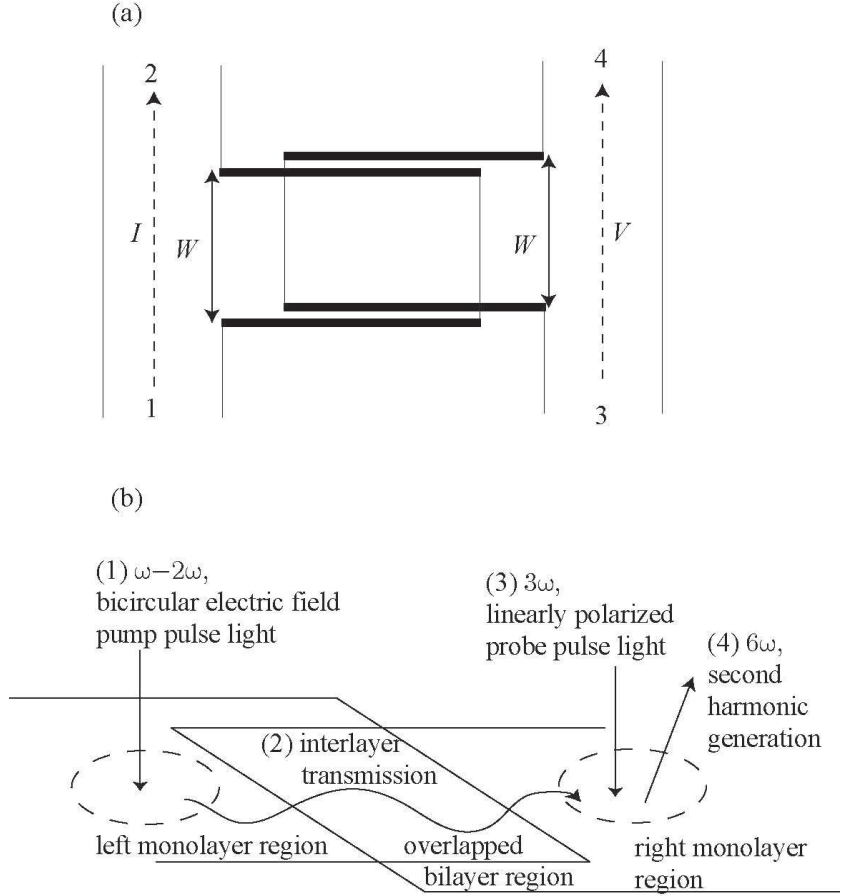}
\caption{Experimental detection of VCR via the bulk states.
(a) Nonlocal resistance $R_{nl}=V/I$, where the current $I$ from contacts 
1 to 2 induces the voltage between contacts 3 and 4.
The central part is similar to the $\downarrow\uparrow$ junction.
The thick lines represent the zigzag edges.
(b) Optical detection with the second harmonic generation.
Dashed circles represent the illuminated areas.
}
\label{fig-experiment}

\end{center}\end{figure}

Figure \ref{fig-experiment}(a)
shows a scheme of an experiment that proves that the bulk states drive the VCR.
The central part is similar to the $\downarrow\uparrow$ junction
with a finite transverse width $W$.
The thick lines represent the zigzag edges.
As the width $W$
increases, the bulk contribution becomes dominant and reduces the difference
between the open and periodic boundary conditions.
The $R_{nl}$ equals $I/(V_3-V_4)$, where $I$ is the charge current
from 1 to 2 probes, and $V_3$ ($V_4$) is the electric potential
at contact 3 (4). It is a VCR signature that the vertical gate
voltage changes the $R_{nl}$ sign.
The top and bottom gates control the parameters $E$ and $\varepsilon$.
In contrast  to the VCR of the zigzag edge states \cite{40.}, the bulk states carry the VC.
The edge is an imperfection of the infinite perfect crystal
and cannot always be controlled.
However, its effects become negligible as the width $W$ increases.
Notably, reducing imperfections is an effective strategy for reproducible outcomes, especially in industrial use.

In electric measurement,
the current terminals, $(V_L, V_R)$ in Fig. 1 and (1, 2) in Fig. 15,
inevitably connect the sample edges and enhance the edge influence
on the current.
In contrast, we can exclude the current terminals in the optical measurements.
The valley polarization can be generated 
by either circularly \cite{a1.,a2.,L-1.,L-5.,L-6.,L-7.,Dixit-1,Dixit-2} or linearly
polarized light \cite{L-3.,L-10.}. We can also measure 
the valley polarization by optical second harmonic (SH) generation \cite{L-8.,L-9.}.
Steps (1)--(4) in Fig. \ref{fig-experiment}(b) illustrate an optical measurement method according to Ref. \cite{Dixit-1} .
(1) The left monolayer region is illuminated with the pump pulse $\omega$ -$2\omega$ bicircular field, where $\omega$ denotes the principal angular frequency.
This illumination induces the spatial gradient of valley concentration, followed by the valley-dependent diffusion of electrons, i.e., VC.
(2) The induced VC is transmitted through the bilayer region with the sign reversed by the vertical gate. Consequently, the right monolayer has the opposite valley polarization compared with the left.
(3) The linearly polarized probe light 
of the angular frequency $3\omega$ is incident to
the right monolayer region and generates the SH.
(4) The valley polarization in the right monolayer is
detectable with the phase of the SH ($6\omega$ angular frequency).
The illuminated areas in steps (1) and (3) are near the bilayer region but are restricted to each monolayer region.
We can also measure the electron transit time in step (2) from the delay time of the probe pulse.

In the experiment described in Ref. \cite{recent-experiment} , the displacement field $D$ times the interlayer distance
$a_{\downarrow\uparrow}$ reaches about 0.5 V with a band gap of 0.14 eV,
where $D$ represents the microscopic electric field felt by the graphene
electrons.
These data are comparable to $(2\varepsilon, \Delta)=(0.7, 0.166)$ eV for the high VCR
$\tilde{g}_{\rm v} =0.8$, suggesting the technical feasibility of the present results.
Since the carriers and surrounding materials have screening effects,
more realistic self-consistent calculations are necessary for the relationship between $eDa_{\downarrow\uparrow}$ and $2\varepsilon$. 
Additionally, the high $D$ might modify the interlayer
transfers $\gamma_1$, $\gamma_3$, and $\gamma_4$. 
The present analytic formulas 
contain no fitting parameter other than the Hamiltonian
elements and thus can be adapted to these advanced calculations
\cite{a12.,(11)}.

\section{Summary and Conclusion}
We discuss the $\downarrow\xi$ junction $(\xi = \downarrow, \uparrow)$, 
 which denotes the double junction
in a series of the left layer $\downarrow$, AB stacking bilayer, and right
layer $\xi$, where $\downarrow$ and $\uparrow$ are the extensions
of the lower and upper layers from the bilayer, respectively.
Using the $\gamma_1$-TB and $\gamma_1\gamma_3\gamma_4$-TB models,
we calculate the transmission rate $T_{\nu',\nu}$
from the left valley $K_{\nu}$
to the right valley $K_{\nu'}$ 
as a function of the energy,
the longitudinal bilayer length,  
and the lateral wave number $\kappa$.
The VCR transmission rate
$g'_{\rm v} (\kappa)=\frac{1}{2}\sum_{\nu}\left [ T_{-\nu,\nu}(\kappa)
- T_{\nu,\nu}(\kappa) \right ]$ averages out at $\tilde{g}_{\rm v}$ per $\kappa$.
The present paper is the first report on
the two VCR origins of the po-G: monolayer--bilayer matching
and interlayer matching.
These two origins refer to `wave function' matching in
position space (not in momentum space) and present an intuitive picture.
Notably, the vertical electric field
enhances the interlayer matching near the bilayer gap edge and provides
the nearly `pure' VCR.
A pure VC changes into an almost pure VC with an inverse sign
and a similar intensity. 
This enhancement occurs only when the electrons are forced to flow vertically, i.e., only in the $\downarrow\uparrow$ junction.
On the other hand, monolayer--bilayer matching is adequate for
the VCR in the $\downarrow\uparrow$ and $\downarrow\downarrow$ 
junctions but weakens under the vertical field.

Using the $\gamma_1$-TB model,
we derive the analytical $T_{\nu',\nu}(0)$
formulas.
Since $\gamma_3$, $\gamma_4$, and nonzero $|\kappa|$
have only minor effects on  $\tilde{g}_{\rm v}$ near the bilayer gap edge,
this formula is effective in the $\tilde{g}_{\rm v}$ peak
of the $\downarrow\uparrow$ junction.
The bilayer length is represented by $(N-2)a/2$ with an integer $N$ and the lattice constant $a$.
Compared with the non-integer $N/3$ case,
the integer $N/3$ case 
shows larger $g'_{\rm v}(0)$ 
and yields Eq. (\ref{3N-T})
that reflects the periodic sublattice localization of Eq. (\ref{def-vec-f}).
The peak $N$ satisfies the condition $\left||k_0|\frac{a}{2}-\frac{2}{3}\pi\right| N \simeq \pi$, where $k_0$ is the bilayer wave number
at the gap edge.

In addition to the valley filter and valley splitter,
the VCR is an essential function in valleytronics but is still in its infancy.
An important contribution is a proposal for the experimental detection
of the VCR driven by the bulk states.
It will open a new field in valleytronics.

\acknowledgments

I thank Gopal Dixit for his helpful suggestions.

\newpage

\appendix

\section{VC Formulas}
The probability conservation 
and the time-reversal symmetry guarantee
\begin{eqnarray} 
T^{\rightarrow}_{\nu,\nu'}(\kappa)=T^{\leftarrow}_{-\nu',-\nu}(-\kappa),
\label{app-T}\\
R^{\leftarrow}_{\nu,\nu'}(\kappa)=R^{\leftarrow}_{-\nu',-\nu}(-\kappa)
\label{app-R-left},\\
R^{\rightarrow}_{\nu,\nu'}(\kappa)=R^{\rightarrow}_{-\nu',-\nu}(-\kappa),
\label{app-R-right}
\end{eqnarray}
where $R$ and $T$ denote the reflection and transmission rates, respectively \cite{a9.}.
The superscript $\leftarrow$ ($\rightarrow$) corresponds
to the incidence from the right (left) monolayer G.
The right (left) subscript indicates the valley
index before (after) the scattering.
Equation (\ref{app-T}) is abbreviated to $T_{\nu,\nu'}(\kappa)$ in
the main text. With this notation, Eq. (\ref{ave-gv}) 
is equivalent to
\begin{equation}
\tilde{g}_{\rm v}=\frac{1}{2M+1}\left[ 
g'_{\rm v}(0)+\sum_{m=1}^M g''_{\rm v}(m\Delta \kappa)
\right],
\end{equation}
where
\begin{equation}
g''_{\rm v}(\kappa)
=\sum_{\nu,\nu'}
\frac{-\nu\nu'}{2}
\left[
T_{\nu',\nu}^{\leftarrow}(\kappa)
+
T^{\rightarrow}_{\nu,\nu'}(\kappa)
\right].
\end{equation}
The inversion of the $\varepsilon$ sign is equivalent to the $\pi$ rotation
around the $y$ axis for the $\downarrow\uparrow$ junction.
Since $g''(\kappa)$ is invariant under this rotation, 
Eq. (\ref{sym1}) is true for
the $\downarrow\uparrow$ junction.
In contrast, the $\downarrow\downarrow$ junction
does not satisfy Eq. (\ref{sym1}) because
this rotation is not equivalent to the $\varepsilon$ sign change.

Using the $\kappa$ average notation 
\begin{equation} 
\tilde{\diamondsuit}=\frac{1}{2M+1}\sum_{m=-M}^M \diamondsuit(m\Delta \kappa),
\end{equation}
we can derive
\begin{eqnarray} 
\tilde{T}^{\rightarrow}_{\nu,\nu'}=\tilde{T}^{\leftarrow}_{-\nu',-\nu},
\label{app-T-2}
\\
\tilde{R}^{\leftarrow}_{\nu,\nu'}=\tilde{R}^{\leftarrow}_{-\nu',-\nu},
\label{app-R-left-2}\\
\tilde{R}^{\rightarrow}_{\nu,\nu'}=\tilde{R}^{\rightarrow}_{-\nu',-\nu}, 
\label{app-R-right-2}
\end{eqnarray}
from Eqs. (\ref{app-T}), (\ref{app-R-left}), and (\ref{app-R-right}).
The VCs in the left (L) and right (R) monolayer regions are
represented by
\begin{equation}
J_{\rm v}^{\rm L}=\sum_{\nu,\nu'}\left[
\left(
\delta_{\nu,\nu'}-\tilde{R}_{\nu',\nu}^{\rightarrow}
\right)\nu'J_\nu^{\rightarrow}
-\tilde{T}_{\nu',\nu}^{\leftarrow}\nu'J_\nu^{\leftarrow}
\right]
\label{Jv-L}
\end{equation}
and
\begin{equation}
J_{\rm v}^{\rm R}=\sum_{\nu,\nu'}\left[
\left(
\tilde{R}_{\nu',\nu}^{\leftarrow}-\delta_{\nu,\nu'}
\right)\nu'J_\nu^{\leftarrow}
+\tilde{T}_{\nu',\nu}^{\rightarrow}\nu'J_\nu^{\rightarrow}
\right]
\label{Jv-R}
\end{equation}
accompanied by the charge current
\begin{eqnarray}
J &=& \sum_{\nu,\nu'}\left[
\left(
\delta_{\nu,\nu'}-\tilde{R}_{\nu',\nu}^{\rightarrow}
\right)J_\nu^{\rightarrow}
-\tilde{T}_{\nu',\nu}^{\leftarrow}J_\nu^{\leftarrow}
\right] \\
&=& 
\sum_{\nu,\nu'}\left[
\left(
\tilde{R}_{\nu',\nu}^{\leftarrow}-\delta_{\nu,\nu'}
\right)J_\nu^{\leftarrow}
+\tilde{T}_{\nu',\nu}^{\rightarrow}J_\nu^{\rightarrow}
\right],
\label{app-J}
\end{eqnarray}
where $J_\nu^{\rightarrow}$ ($J_\nu^{\leftarrow}$ ) 
denotes the non-negative incidence flow from the $K_\nu$ valley of region L (R).

Typical VCR occurs under two incidence conditions (i)
$J_\nu^{\rightarrow} =1, J_{-\nu}^{\rightarrow}
=J_+^{\leftarrow} = J_-^{\leftarrow} =0$ 
and (ii) $J_\nu^{\rightarrow}=J_\nu^{\leftarrow} =1, J_{-\nu}^{\rightarrow}=J_{-\nu}^{\leftarrow} =0$.
Under condition (i), $J_{\rm v}^{\rm R} 
=\tilde{T}_{+,\nu}^\rightarrow-\tilde{T}_{-,\nu}^\rightarrow$,
\begin{equation}
J_{\rm v}^{\rm L} 
=\nu 
(1-\tilde{R}_{\nu,\nu}^\rightarrow+\tilde{R}_{-\nu,\nu}^\rightarrow),
\end{equation}
and $J
=\tilde{T}_{+,\nu}^\rightarrow+\tilde{T}_{-\,\nu}^\rightarrow$.
Under condition (ii), 
\begin{equation}
J_{\rm v}^{\rm L}
=\nu (
\tilde{R}_{-,+}^\rightarrow+\tilde{R}_{+,-}^\rightarrow
+
\tilde{T}_{-,+}^\rightarrow+\tilde{T}_{+,-}^\rightarrow
),
\end{equation}
\begin{equation}
J_{\rm v}^{\rm R}
=-\nu (
\tilde{R}_{-,+}^\leftarrow+\tilde{R}_{+,-}^\leftarrow
+
\tilde{T}_{-,+}^\rightarrow+\tilde{T}_{+,-}^\rightarrow
),
\end{equation}
and $J
=\tilde{T}_{\nu,\nu}^\rightarrow-\tilde{T}_{-\nu,-\nu}^\rightarrow$.
When $\tilde{g}_{\rm v} \simeq 1$, the reflection rates $R_{\nu,\nu'}$ and 
intravalley transmission rates $T_{\pm,\pm}$ are near zero followed by the VCR $J_{\rm v}^{\rm L}\simeq -J_{\rm v}^{\rm R}$.
In  particular, the VCR of condition (ii) corresponds to the reversal of pure VC as
$|J_{\rm v}| \simeq 2 \gg |J|$.
In the calculation with condition (ii) above, we use
Eqs. (\ref{app-T-2}), (\ref{app-R-left-2}), and (\ref{app-R-right-2}), 
and the probability conservation
\begin{equation}
\sum_{\nu'} \tilde{T}_{\nu',\nu}^\leftarrow+\tilde{R}_{\nu',\nu}^\leftarrow
=\sum_{\nu'} \tilde{T}_{\nu',\nu}^\rightarrow+\tilde{R}_{\nu',\nu}^\rightarrow=1.
\end{equation}

\section{Explicit Expressions of the Second Approximation} 
Except for the last paragraph, in the present Appendix, we  discuss the
$\downarrow\uparrow$ junction.
In the case of the $\downarrow\uparrow$ junction,
\begin{equation}
G_{\;^B_A}
=
\frac{\gamma_1(|\theta_\mp|s_- -|\theta_\pm|s_+)}
{\sqrt{3}|\gamma_0|(\theta_+^2-\theta_-^2)},
\label{G}
\end{equation}

\begin{equation}
{\rm Re} [F_A]
=\frac{3\gamma_0^2|\theta_+\theta_-|}{\varepsilon^2-E^2}
(Z+c_+c_-) +s_+s_-,
\label{ReFA}
\end{equation}

\begin{equation}
{\rm Re} [F_B] 
=
-Z -c_+c_- + \frac{3\gamma_0^2|\theta_+\theta_-|}{E^2-\varepsilon^2}s_+s_-,
\label{ReFB}
\end{equation}

\begin{eqnarray}
{\rm Im} [ F_{\;_A^B}]
&=
\frac{\sqrt{3}|E\gamma_0|}{E^2-\varepsilon^2}
&
\left[ 
|\theta_\pm| s_+c_-+ |\theta_\mp| s_-c_+ \right.
\nonumber \\
&& \left. \mp
\frac{4\varepsilon^2(|\theta_\pm| s_+c_-- |\theta_\mp| s_-c_+)}
{3\gamma_0^2(\theta_+^2-\theta_-^2)}
\right],
\label{ImF} 
\end{eqnarray}
where
\begin{eqnarray}
Z
&=
\frac{2\gamma_1^2(E^2-\varepsilon^2)}
{9\gamma_0^4(\theta_+^2-\theta_-^2)^2}
&
\left[
\left(
1-\frac{\theta_+^2+\theta_-^2}{2|\theta_+\theta_-|}
\right)s_+s_- \right.
\nonumber \\
&& \left. + 2\sin^2\left(\frac{|\theta_+|-|\theta_-|}{2}N\right)
\right],
\label{Z}
\end{eqnarray}

\begin{equation}
c_\pm=\cos(\theta_\pm N),\;\;
s_\pm=\sin(|\theta_\pm|N).
\label{app-cs}
\end{equation}

When $|E|$ approaches $\Delta$, 
$|\theta_\pm|$ converges to $\theta_0$, followed by the limits 
\begin{equation}
\lim_{|\theta_\pm| \rightarrow \theta_0 }
G_{\;_A^B}
= - \sqrt{\frac{1+q}{q(2+q)}}\;(
\phi_0\cos\phi_0\pm \sin\phi_0),
\label{G-du-2} 
\end{equation}

\begin{eqnarray}
\lim_{|\theta_\pm| \rightarrow \theta_0 }
F_B &= & 
-1-\frac{2}{q} 
+\frac{1+q}{(2+q)^2}(\phi_0^2-\sin^2\phi_0)
\nonumber \\
& & +\frac{2}{q}\cos^2\phi_0-i\frac{\sin (2\phi_0)}{q\sqrt{2+q}}+2i\frac{1+q}{q\sqrt{2+q}}\phi_0,
\label{F+2} 
\end{eqnarray}
and
\begin{eqnarray}
\lim_{|\theta_\pm| \rightarrow \theta_0 }
F_A &=& 
1 
+\frac{1+q}{q(2+q)}(\sin^2\phi_0-\phi_0^2)
\nonumber \\
& & 
+\frac{2}{q}\cos^2\phi_0
-i\frac{\sin(2\phi_0)}{q\sqrt{2+q}}-2i\frac{1+q}{q\sqrt{2+q}}\phi_0,
\label{F-2} 
\end{eqnarray}
where $\phi_0=\theta_0 N$. 
Under condition (\ref{cond-theta0}), 
the last terms become dominant
in Eqs. (\ref{F+2}) and (\ref{F-2}).
Equation (\ref{3N-T-2})
originates from
these terms and Eq. (\ref{G-du-2}).

\begin{figure}
\begin{center}
\includegraphics[width=0.5\linewidth]{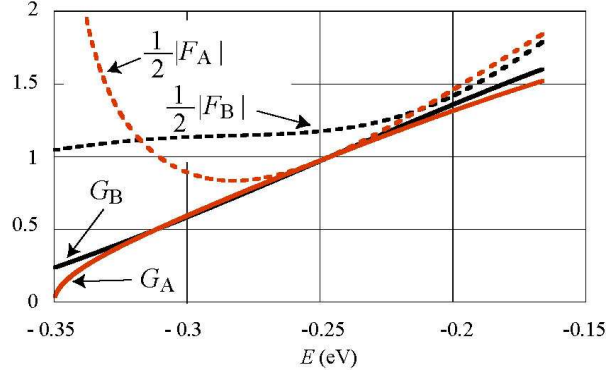}
\caption{(Color online) $|F|/2$ and $G$ calculated using
Eqs. (\ref{G}) (\ref{ReFA}), (\ref{ReFB}), and (\ref{ImF})
in the case of $\varepsilon= 0.35$ eV and $N=45$.
}
\end{center}
\label{fig-appendix}
\end{figure}

Figure \ref{fig-band} demonstrates that $|\theta_+| + |\theta_-| \simeq 2\theta_0$ 
when $|E|$ is near $\Delta$.
This leads to the approximation
\begin{equation} 
N|\theta_{\pm}| \simeq \pi(1 \pm \varphi),
\label{phi}
\end{equation} 
where $N \simeq \pi/\theta_0$ (Eq. (\ref{Np})).
When $|\varphi|< \frac{1}{3}$,
\begin{equation} 
\Delta < |E| < \frac{\Delta}{3}\sqrt{11+q},
\label{Ep-range}
\end{equation} 
as is confirmed below.
The smallness of $|\sin(\theta_0 N)|$
is necessary for the derivation of
the following approximate formulas from Eq. (\ref{phi}).
Since $N/3$ is an integer, 
$\theta_0 N$ equals $3\theta_0$ times an integer, 
and thus the minimum $|\sin(\theta_0 N)|$
can reach $\sin(3 \theta_0 /2)$.
The condition $q<8$ is equivalent to the condition
$\sin(3\theta_0/2) < 0.16$.
First,  we discuss the case $1<q< 8$.
The other case, $q< 1$, is discussed later.

When $|E|$ reaches Eq. (\ref{Ep}), Eq. (\ref{q-theta}) 
produces
\begin{equation}
\theta_+^2+\theta_-^2=\frac{10q(2+q)\gamma_1^2}{54(1+q)\gamma_0^2}.
\label{P-peak2}
\end{equation}
This approximates
\begin{equation}
\theta_+^2+\theta_-^2
\simeq \left[
10-\frac{1}{(1+q)^2}
\right] \frac{q(2+q)\gamma_1^2}{54(1+q)\gamma_0^2}
\label{P-peak}
\end{equation}
because $\frac{1}{(1+q)^2} \ll 10$.
On the other hand, Eq. (\ref{q-theta})
guarantees an identity 
\begin{equation}
\theta_+^2+\theta_-^2
=\frac{3(\theta_+^2-\theta_-^2)^2\gamma_0^2}{2(1+q)\gamma_1^2}
+\frac{q(2+q)\gamma_1^2}{6(1+q)\gamma_0^2}.
\label{identity}
\end{equation}
Equations (\ref{P-peak}) and (\ref{identity}) result in
\begin{equation}
\theta_+^2-\theta_-^2
\simeq \frac{q(2+q)\gamma_1^2}{9(1+q)\gamma_0^2}.
\label{Q-peak}
\end{equation}
Equations (\ref{P-peak2}) and (\ref{Q-peak})
satisfy condition $|\theta_+| \simeq 2|\theta_-|$.
When this condition and Eq. (\ref{phi})
hold, $\varphi \simeq \frac{1}{3}$.
When Eq. (\ref{phi}) is effective, Eqs. (\ref{G}),
(\ref{ReFA}), (\ref{ReFB}), and (\ref{ImF})
are approximated by 
\begin{eqnarray}
G_B & \simeq & G_A \label{GB=GA} \\
& \simeq & \frac{\sin(\pi\varphi)}{\varphi}\sqrt{\frac{1+q}{q(2+q)}},
\label{phi-G} 
\end{eqnarray}
\begin{eqnarray}
{\rm Re} [F_A] & \simeq & \cos(2\pi\varphi)
-\frac{1+q}{q(2+q)}\frac{\sin^2(\pi\varphi)}{\varphi^2} \nonumber \\
& & +\frac{2}{q}\cos^2(\pi\varphi),
\label{phi-FA} 
\end{eqnarray}
\begin{eqnarray}
{\rm Re} [F_B] & \simeq & - {\rm Re} [F_A]
-\frac{2(1+q)\sin^2(\pi\varphi)}{q(2+q)^2\varphi^2} \nonumber \\
& & +\frac{2}{q},
\label{phi-FB} 
\end{eqnarray}
and 
\begin{eqnarray}
{\rm Im} [F_B]
&\simeq & -{\rm Im} [F_A] \label{phi-ImF2} \\
&\simeq & 
\frac{1+q}{q}\sqrt{\varphi^2+\frac{1}{2+q}}
\frac{\sin(2\pi\varphi)}{\varphi}.
\label{phi-ImF} 
\end{eqnarray}
Equations (\ref{phi-FB}) and (\ref{phi-ImF2})
suggest that $F_{\rm B} \simeq -F_{\rm A}$.
Equations (\ref{ReFA}), (\ref{ReFB}), and (\ref{ImF}) certainly satisfy the
inequality $\left|{\rm Re} [F_A+F_B]/{\rm Im} [F]\right| < 0.33$
when $1 < q < 8$ and $|\varphi| <\frac{1}{3}$.
The relation $F_{\rm B} \simeq -F_{\rm A}$
and Eq. (\ref{GB=GA})
cause the constructive VCR interference between the 
$\vec{f}_A$ and $\vec{f}_{B}$
modes as $g'_{\rm v}(0) \simeq 4|G_A/F_A|^2$.
Figure \ref{fig-appendix} displays 
$|F|/2$ and $G$ 
in the case of $\varepsilon= 0.35$ eV $(q=3.45)$ and $N=45$ 
for the negative inner $E$ region $( -\varepsilon < E < -\Delta$),  
where $\pi/\theta_0 = 43.8$, $\Delta=0.166$ eV, 
and $\frac{\Delta}{3}\sqrt{11+q} =$ 0.21 eV.
When $\varphi$ increases from 0 to $1/3$,
$|F|/2$ approaches $G$, and the increase in
$4G^2/|F|^2$ follows.
On the other hand, Eqs. (\ref{G}) and (\ref{ReFA}) prove that 
$|G_A/F_A|$ converges to zero at $|E|=|\varepsilon|$, followed
by the disappearance  of $g_{\rm v}'(0)$.
These results indicate that Eqs. (\ref{Np}) and (\ref{Ep})
correspond to the $g'_{\rm v}(0)$ peak.
At the same time, Fig. B-1 shows 
that the variation of $|G/F|$ is minimal in the $E$ range (\ref{Ep-range}).
We have discussed above the case where $1<q<8$.
As $q$ decreases from 1, on the other hand, the inner $E$ region shrinks, and
thus the variation of $|G/F|$ in the inner $E$ region becomes small.
In summary, the peak $g'_{\rm v}(0)$
in conditions (\ref{Np}) and (\ref{Ep}) 
is similar to $q/(1+q)$ calculated using Eqs. (\ref{3N-T-2})
and (\ref{Np}).
When $|E| \gg \sqrt{4\varepsilon^2+\gamma_1^2}$, on the other hand, we derive 
Eq. (\ref{3N-T-3}) using
$3\gamma_0^2\theta_\pm^2\simeq E^2\pm|E|\sqrt{4\varepsilon^2+\gamma_1^2}$, 
$Z \simeq \frac{1}{1+q}\sin^2\left(\frac{|\theta_+|-|\theta_-|}{2}N\right)$, 
${\rm Im}(F_A) \simeq {\rm Im}(F_B) 
\simeq \sin\left((|\theta_+|+|\theta_-|)N\right)$, and $3\gamma_0^2|\theta_+\theta_-|/(E^2-\varepsilon^2) \simeq 1$.

In the case of the $\downarrow\downarrow$ junction,
\begin{eqnarray}
{\rm Im} [F_{\;_A^B}]
&=
\frac{\sqrt{3}|\gamma_0|}{(E+\varepsilon)}
&
\left[
|\theta_\pm|s_+c_-+|\theta_\mp|s_-c_+
\right.
\nonumber \\
&& \left.+\frac{4\varepsilon E(|\theta_\mp|s_-c_+-|\theta_\pm|s_+c_-)}
{
3\gamma_0^2(\theta_+^2-\theta_-^2)
}
\right
],
\end{eqnarray}
\begin{eqnarray}
G_{\;_A^B}
&=
\frac{\sqrt{3}|\gamma_0|
}{2(E+\varepsilon)}
&
\left[ 
\frac{
4\varepsilon E
(
|\theta_\mp|s_--|\theta_\pm|s_+
)
}
{
3\gamma_0^2(\theta_+^2-\theta_-^2)
}
\right.
\nonumber \\
&&
\left.
+|\theta_\mp|s_-+|\theta_\pm|s_+
\right],
\end{eqnarray}
\begin{equation}
{\rm Re} [F_B]
=Z +
\left[
\frac{3\gamma_0^2|\theta_+\theta_-|}{(E+\varepsilon)^2}
+
\frac{(E-\varepsilon)^2}{3\gamma_0^2|\theta_+\theta_-|}
\right]s_+s_-,
\end{equation}
and
\begin{equation}
{\rm Re} [F_A]
=\frac{3\gamma_0^2|\theta_+\theta_-|}{(E+\varepsilon)^2}Z
+2s_+s_-.
\end{equation}


\begin{thebibliography}{99}
\bibitem{note}
The present paper is different from Ref. 67
in the definitions of $\varepsilon$ and $E$.
$(E, 2\varepsilon, E+\varepsilon)=
(\widetilde{E}, \varepsilon, E)_{\rm Ref. 67}$.
\end{thebibliography}
\end{document}